\documentclass[aps,prb,twocolumn,superscriptaddress]{revtex4-2}
\usepackage{color}
\usepackage{graphicx}
\usepackage{amsmath,amsfonts,amssymb}
\usepackage{microtype}
\usepackage{float}
\usepackage{siunitx}
\usepackage{subfigure}
\usepackage{bm}
\usepackage[colorlinks=true, letterpaper=true, pdfstartview=FitV, linkcolor=red, citecolor=blue, urlcolor=blue]{hyperref}

\begin{document}
\title{Majorana Flat Bands in the Vortex Line of Superconducting Weyl Semimetals}

\author{Zhicheng Zhang}\email{zhicheng_zhang1995@pku.edu.cn}
\affiliation{International Center for Quantum Materials and School of Physics, Peking University, Beijing 100871, China}
\affiliation{Hefei National Laboratory, Hefei 230088, China}

\author{Kou-Han Ma}
\affiliation{International Center for Quantum Materials and School of Physics, Peking University, Beijing 100871, China}
\affiliation{Hefei National Laboratory, Hefei 230088, China}

\begin{abstract}
We report the emergence of Majorana flat bands (MFBs) in the vortex line of superconducting (SC) time-reversal-symmetry-breaking Weyl semimetals. By considering a Weyl semimetal as a stack of Chern insulators with varying Chern numbers along one ($z$) direction, we decompose the vortex bound states of SC Weyl semimetals into those of $k_{z}$-resolved SC Chern insulators. Through analytical and numerical calculations of  the topological phase diagram of the SC Chern insulators, we explain the appearance of MFBs and determine the exact boundaries of them. Notably, the tuning of  chemical potential or pairing strength results in the MFBs along the entire $k_{z}$ axis. To characterize the MFBs,  we propose a $k_{z}$-resolved $Z_{2}$ Chern-Simons invariant as the topological indicator. Finally, we take an attractive Hubbard interaction into consideration, and the aforementioned SC Weyl semimetal with BCS pairing can be realized under appropriate parameters.  
\end{abstract} 


\maketitle

\section{\label{s:intro}Introduction}

The discovery of emergent $p$-wave superconductivity in the $s$-wave superconductor-TI heterostructure \cite{PhysRevLett.100.096407,RevModPhys.82.3045,RevModPhys.83.1057} has spurred significant interest  in the vortex bound states (VBSs) of superconducting (SC) topological materials \cite{alicea2012new,zhang2009topological,PhysRevLett.100.096407,PhysRevLett.105.097001,PhysRevB.84.144507,PhysRevLett.109.237009,hor2011superconductivity,zhang2011pressure,PhysRevLett.107.097001,PhysRevLett.114.017001,PhysRevLett.116.257003,PhysRevLett.125.037001,zhang2022bulkvortex,hu2023topological,PhysRevLett.108.140405,BJYang,PhysRevB.88.125427,PhysRevB.85.195320,liu2014stable,he2016pressure,PhysRevX.11.021065,zhou2021bulk,PhysRevLett.130.046402,PhysRevB.105.014509,PhysRevLett.123.027003,PhysRevLett.124.257001,PhysRevLett.119.047001,PhysRevLett.115.187001,fei2018band, kobayashi2023crystal, PhysRevB.107.115127} and topological unconventional superconductors \cite{PhysRevLett.117.047001,PhysRevB.92.115119,PhysRevB.93.115129,qin2019topological,PhysRevLett.129.277001,PhysRevB.103.L140502,wang2018evidence,zhang2018observation,kong2019half,PhysRevB.101.020504,PhysRevLett.122.207001}.
Among them, the VBSs of SC Weyl semimetals show many exotic and intriguing properties \cite{PhysRevLett.107.127205,PhysRevLett.107.186806,PhysRevB.83.205101,nomani2023intrinsic,deng2022pressure,kononov2021superconductivity,PhysRevB.105.174502,tu2022superconductivity,PhysRevMaterials.5.084201,van2020two,PhysRevX.5.011029,huang2015weyl,li2017concurrence,PhysRevB.107.144512,PhysRevB.97.020501,PhysRevLett.118.207002,PhysRevLett.124.257001,zhang2021gapless,PhysRevLett.127.187002,PhysRevLett.130.156402}. Particularly, for a SC  time-reversal-symmetry-breaking Weyl semimetal in the pair density wave state, there are gapless chiral Majorana modes in the vortex line protected by an emergent second Chern number \cite{PhysRevLett.118.207002}. In contrast, a SC time-reversal-symmetric Weyl semimetal has Majorana zero modes (MZMs) at the ends of the vortex line, and topological invariants are also proposed to characterize such MZMs \cite{PhysRevLett.127.187002,PhysRevLett.124.257001}.  Experimental technologies, such as point contact \cite{wang2016observation} and selective ion sputtering, \cite{bachmann2017inducing} have also been developed to induce superconductivity into topological semimetals, which motivates further theoretical study about the VBSs of SC Weyl semimetals.

Majorana flat bands (MFBs) are zero energy flat bands in the BdG spectra of a superconductor. They were found in the study of VBSs of 3D spinless chiral superconductor of the ${}^{3}\rm{He}$-A type, where the boundaries of the flat bands are determined by the locations of the two gapless points of the bulk spectra \cite{volovik2011flat}. Subsequent studies revealed  that after considering the in-plane magnetic field, a chiral $p_{x}\pm ip_{y}$-wave superconductor becomes a gapless superconductor and host MFBs \cite{PhysRevB.88.060504}. Further work reported  the existence of MFBs in noncentrosymmetric superconductors \cite{PhysRevB.105.184501,PhysRevB.95.134507}, $d$-wave superconductors \cite{yuan2014probing}, and SC Weyl semimetals \cite{PhysRevB.93.201105}. A recent study has also uncovered the hinge-localized MFBs in 
\begin{figure}[tbph]
	\centering
	\includegraphics[scale=0.287]{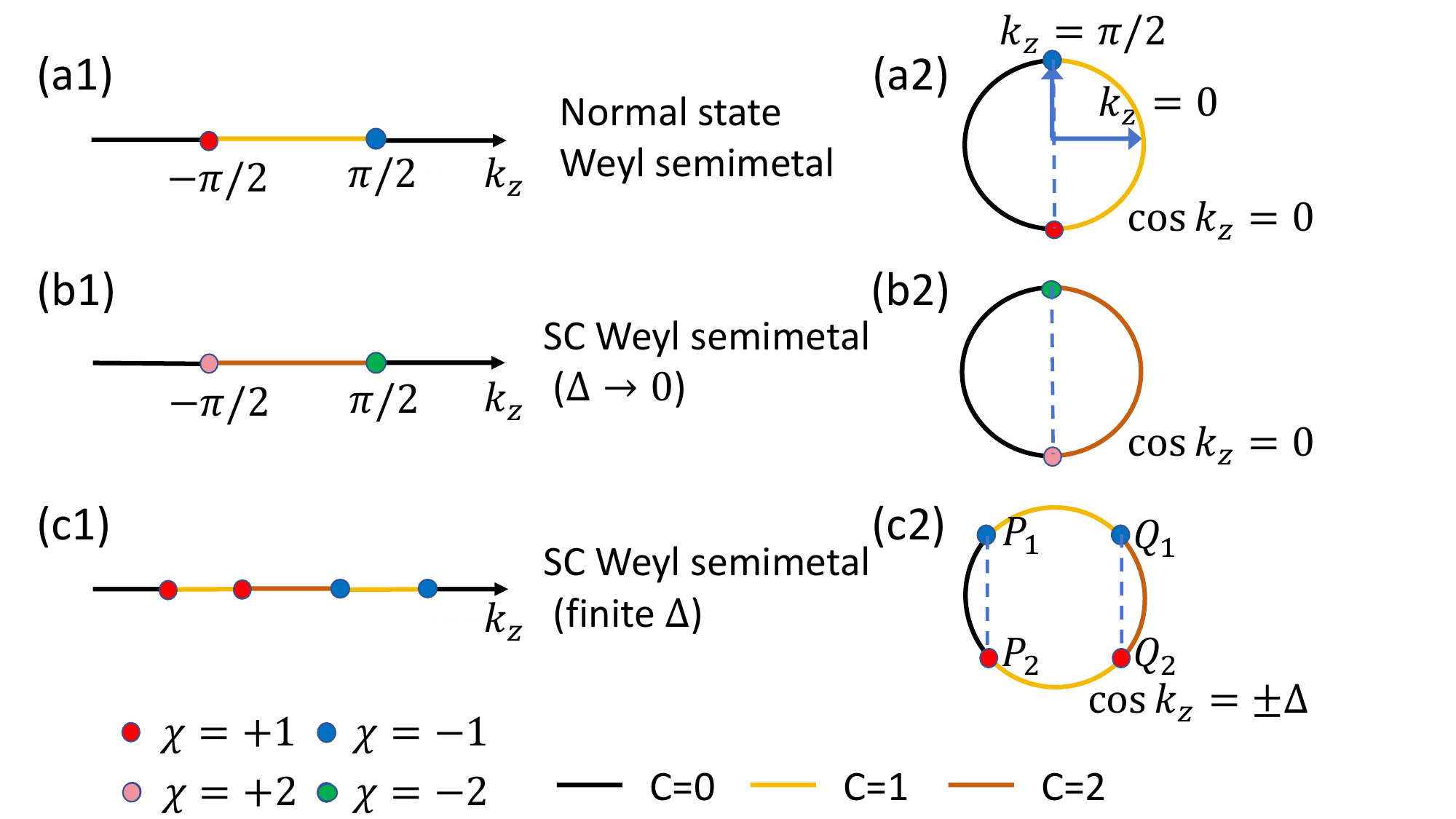}
	\caption{The evolution of bulk gapless points of the Weyl semimetal and the explanation for the appearance of MFBs ($\mu=0$ case). (a1) The Weyl points of the normal state Weyl semimetal are located at $k_{z}=\pi/2$ and $k_{z}=-\pi/2$. The two-dimensional insulators between the two Weyl points have Chern number $C=1$, which is indicated by the yellow line. (a2) takes the periodic boundary condition in $k_{z}$ axis, and the $k_{z}$ axis now becomes a unit circle. It shows that for the model of Weyl semimetal we consider, the bulk gapless points $k_{z}=\pm \pi/2$ are given by the condition $\cos k_{z}=0$. (b1) The bulk gapless points of SC Weyl semimetal (BdG Hamiltonian) with $\Delta=0$ are also located at $k_{z}=\pm\pi/2$, but the Chern numbers of the two-dimensional insulators at each $k_{z}$ have doubled. (b2) shows that the bulk gapless points are also given by $\cos k_{z}=0$. (c1) There are four bulk gapless points for SC Weyl semimetals with finite $\Delta$, and $C=1$ regions appear between $C=0$ and $C=2$ regions. (c2) shows that the four gapless points are given by the condition $\cos k_{z}=\pm \Delta$.}  
    \label{F1}
\end{figure} 
type-$\rm{\uppercase\expandafter{\romannumeral2}}$ Dirac semimetals with unconventional pairing \cite{xie2024hinge}.   Despite the intensive exploration of MFBs in SC topological materials, the MFBs in the vortex line of SC Weyl semimetals remain underexplored \cite{sun2019vortices,PhysRevB.86.054504,PhysRevLett.118.207002}. 

Motivated by these considerations, we study the VBSs of SC time-reversal-symmetry-breaking Weyl semimetals. We find that there are MFBs in the vortex line (along $z$ direction), and tuning the chemical potential $\mu$ or the pairing strength  $\Delta$ leads to the MFBs along the whole $k_{z}$ axis. As illustrated in Fig. \ref{F1}(a1), a Weyl semimetal can be regarded as Chern insulators with varying Chern numbers stacked along $z$ direction, so we understand the VBSs of SC Weyl semimetals by decomposing them into $k_{z}$-dependent SC Chern insulators. We calculate the topological phase diagram of the SC Chern insulators both analytically and numerically. A fixed $k_{z}$ in the model of SC Weyl semimetal gives a SC Chern insulator, so the SC Weyl semimetal corresponds to a line in the topological phase diagram. Thus, we can determine the BdG Chern number of the two-dimensional models obtained by fixing $k_{z}$ in the SC Weyl semimetal. There are certain ranges of $k_{z}$ where the BdG Chern number is $C=1$, corresponding to the yellow lines with $C=1$ in Fig. \ref{F1}(c1).  Since there is a MZM in the vortex core of a chiral topological superconductor with odd BdG Chern number \cite{PhysRevB.82.184516}, the MZMs in certain ranges of $k_{z}$ with $C=1$ will constitute the MFBs. We thus explain the appearance of MFBs and determine the exact boundaries of them. Notably, discrepancies arise between the numerically calculated  MFB regions and the analytical phase boundaries  due to the hybridization between the vortex-core and the edge-localized MZMs. We also propose a $k_{z}$-dependent $Z_{2}$ Chern-Simons invariant to characterize the MFBs in the vortex line. Finally, adding the attractive Hubbard interaction into the model of Weyl semimetal, we employ the  mean-field methods and obtain  the aforementioned SC Weyl semimetals with BCS pairing and the MFBs in the vortex line under appropriate parameters.  

\section{\label{s:intro}Models of Chern Insulators and Weyl Semimetals}
We start with the model of a Chern insulator, which is the famous Qi-Wu-Zhang model \cite{PhysRevB.74.085308}.
\begin{equation}
	\begin{aligned}
		H_{\rm{CI}}(\bm{k})&=[m+2B(2-\cos k_{x}-\cos k_{y})]\sigma_{3}\\
		& +A \sin k_{x} \sigma_{1}+A\sin k_{y}\sigma_{2}
	\end{aligned}
	\label{E1}
\end{equation}
For the sake of concreteness, we set the parameters as $A=1$ and $B=-0.5$ throughout this article. The properties of such a Chern insulator are well
known. The bulk gap closes at $m=0$, $m=2$, and $m=4$, which are the three topological phase transition points. We can calculate the Chern number of this Chern insulator and obtain $C=0$ for $m<0$ or $m>4$, $C=1$ for $0<m<2$, and $C=-1$ for $2<m<4$. 

\subsection{The Normal State Properties of Weyl Semimetal}
Then we write down the corresponding model of  a 3D time reversal symmetry breaking Weyl semimetal.
\begin{equation}
	\begin{aligned}
	H_{\rm{WSM}}(\bm{k})&=[2t_{z}\cos k_{z}+2B_{W}(2-\cos k_{x}-\cos k_{y})]\sigma_{3}\\
	& +A_{W} \sin k_{x} \sigma_{1}+A_{W}\sin k_{y}\sigma_{2}
	\end{aligned}
 \label{E2}
\end{equation}
We have chosen the names of the parameters in the above Hamiltonian deliberately, so that parameters $A_{W}$ and $B_{W}$  in Eq. (\ref{E2}) correspond to $A$ and $B$ in Eq. (\ref{E1}), respectively. What's more, $2t_{z}\cos k_{z}$ in Eq. (\ref{E2}) corresponds to $m$ in Eq. (\ref{E1}). Since we have set $A=1$ and $B=-0.5$ in Eq. (\ref{E1}), we set the parameters as $A_{W}=1$, $B_{W}=-0.5$, and $t_{z}$ is set to be $0.5$ in the model of the Weyl semimetal Eq. (\ref{E2}). Under this set of parameters, it is easy to find that the locations of the Weyl points are determined by the gap closing condition $\cos k_{z}=0$. Thus, as shown in Fig. \ref{F1}(a1), the model Eq. (\ref{E2}) describes a Weyl semimetal with two Weyl points located at $(0, 0, \pm\pi/2)$ on the $k_{z}$ axis. It is well known that a time reversal symmetric Weyl semimetal has at least four Weyl points \cite{RevModPhys.90.015001}, so we claim that the Weyl semimetal Eq. (\ref{E2}) breaks time reversal symmetry. Fig. \ref{F1}(a2) takes the periodic boundary condition in $k_{z}$ axis, and the $k_{z}$ axis now becomes a unit circle. The coordinate of a point on this unit circle is $(\cos k_{z},  \sin k_{z})$. Thus, the blue dashed line shows that the gap closing condition $\cos k_{z}=0$ leads to the two Weyl points located at $k_{z}=\pi/2$ and $k_{z}=-\pi/2$. As indicated in the legend, the red and blue points are Weyl points with opposite chirality. 

If we fix the value of $k_{z}$ in Eq. (\ref{E2}), we will obtain a 2D insulator ($k_{z}\neq \pm\pi/2$), in which we can define the Chern number. For $-\pi/2<k_{z}<\pi/2$, we have $2t_{z}\cos k_{z}\in (0,1)$. Because $2t_{z}\cos k_{z}$ corresponds to $m$ in the model of Chern insulator Eq. (\ref{E1}), and $C=1$ for $0<m<2$, we conclude that $C=1$ for insulators at $k_{z}$ with $-\pi/2<k_{z}<\pi/2$ in the 3D Weyl semimetal, which is represented by the yellow line in Fig. \ref{F1}(a1). In contrast, for $-\pi<k_{z}<-\pi/2$ or $\pi/2<k_{z}<\pi$, we have $2t_{z}\cos k_{z}\in (-1,0)$. Because $C=0$ for $m<0$ in the model of Chern insulator Eq. (\ref{E1}), we have $C=0$ for insulators at $k_{z}$ with $-\pi<k_{z}<-\pi/2$ or $\pi/2<k_{z}<\pi$ in the Weyl semimetal, which corresponds to the black lines in Fig. \ref{F1}(a). 

\subsection{The Properties of Superconducting Weyl Semimetals}
Above discussions focus on the properties of normal state Weyl semimetals. To study the VBSs of SC Weyl semimetals, we have to consider the pairing terms. Thus, we consider the following BdG Hamiltonian describing a SC Weyl semimetal with BCS pairing under the basis $\Psi^{\dagger}_{\bm{k}}=(c^{\dagger}_{\bm{k}\uparrow}, c^{\dagger}_{\bm{k}\downarrow}, c_{\bm{-k}\uparrow}, c_{\bm{k}\uparrow})$. 

\begin{equation}
H_{\rm{BdG}}(\bm{k})=\left( \begin{matrix} H_{N}(\bm{k})-\mu & \tilde{\Delta}\\  \tilde{\Delta}^{\dagger} & \mu-H^{*}_{N}(-\bm{k}) \end{matrix}\right).
\label{E3}
\end{equation}
Here $H_{N}(\bm{k})$ refers to the normal state Hamiltonian of 2D Chern insulators or 3D Weyl semimetals, and $\tilde{\Delta}=\Delta i\sigma_{2}$.

For simplicity, we first consider the case with $\mu=0$ and $\Delta=0$. When $\Delta=0$, $H_{\rm{BdG}}$ can be decoupled into two $2\times2$ blocks, and for the two-dimensional model at a fixed $k_{z}$, the Chern numbers of the two blocks are the same. As a result, the Chern number of the two-dimensional insulator at each $k_{z}$ is twice that of the normal state Weyl semimetal, which is shown in Fig. \ref{F1}(b1). On the other hand, Fig. \ref{F1}(b2) schematically shows that both the gap closing conditions of the two blocks are $\cos k_{z}=0$, and there are doubly degenerate Weyl points at $k_{z}=\pi/2$ and $k_{z}=-\pi/2$.  

It turns out that a finite $\Delta$ breaks the two-fold degeneracy of Weyl points at $k_{z}=\pi/2$ and $k_{z}=-\pi/2$. In the following sections, we will calculate the topological  phase diagram of SC Chern insulators and consider the SC Weyl semimetal as a line in the phase diagram. In this process, we show that the gap closing condition for the SC Weyl semimetal Eq. (\ref{E3}) with finite $\Delta$ and $\mu=0$ is $\cos k_{z}=\pm \Delta$. As shown in Fig. \ref{F1}(c2), this gap closing condition leads to four gapless points on the $k_{z}$ axis, and the $C=1$ region (the yellow lines) appear between $C=0$ and $C=2$ regions. Because there is a MZM trapped in the vortex core of a chiral topological superconductor with odd BdG Chern number, when we calculate the VBSs for two-dimensional models with $k_{z}$ in the $C=1$ region, there are MZMs in the VBSs. These MZMs will constitute the MFBs, and the boundaries of the MFBs are also the boundaries of the $C=1$ region, which are given by the four gapless points in Fig. \ref{F1}(c1). This is the general picture illustrating why there are MFBs in the VBSs of SC Weyl semimetals breaking time reversal symmetry. 
The behavior of VBSs of SC Weyl semimetals is more rich in $\mu\neq0$ case, and we will discuss the $\mu\neq0$ case in the following sections.

To calculate the VBSs of SC Weyl semimetals, we have to add a $\pi$-flux inserted into the superconductor (along $z$ direction).  Such a $\pi$-flux line can be described by attaching the phase $e^{i\theta}$ ($\theta=\arctan y/x$) to the pairing potential. Obviously, the vortex line breaks the translation symmetry in $x$ and $y$ directions but preserves the translation symmetry in $z$ direction. Therefore, we choose the open boundary conditions in $x$ and $y$ directions and calculate the VBSs for each $k_{z}$. Besides, we assume that the pairing potential at the vortex core is $\Delta=0$, and $\Delta(\bm{r})=\Delta e^{i\theta}$ on other lattice sites. 

\section{The Topological Phase Diagram of Superconducting Chern Insulators}
\begin{figure}[tbph]
	\centering
	\includegraphics[scale=0.292]{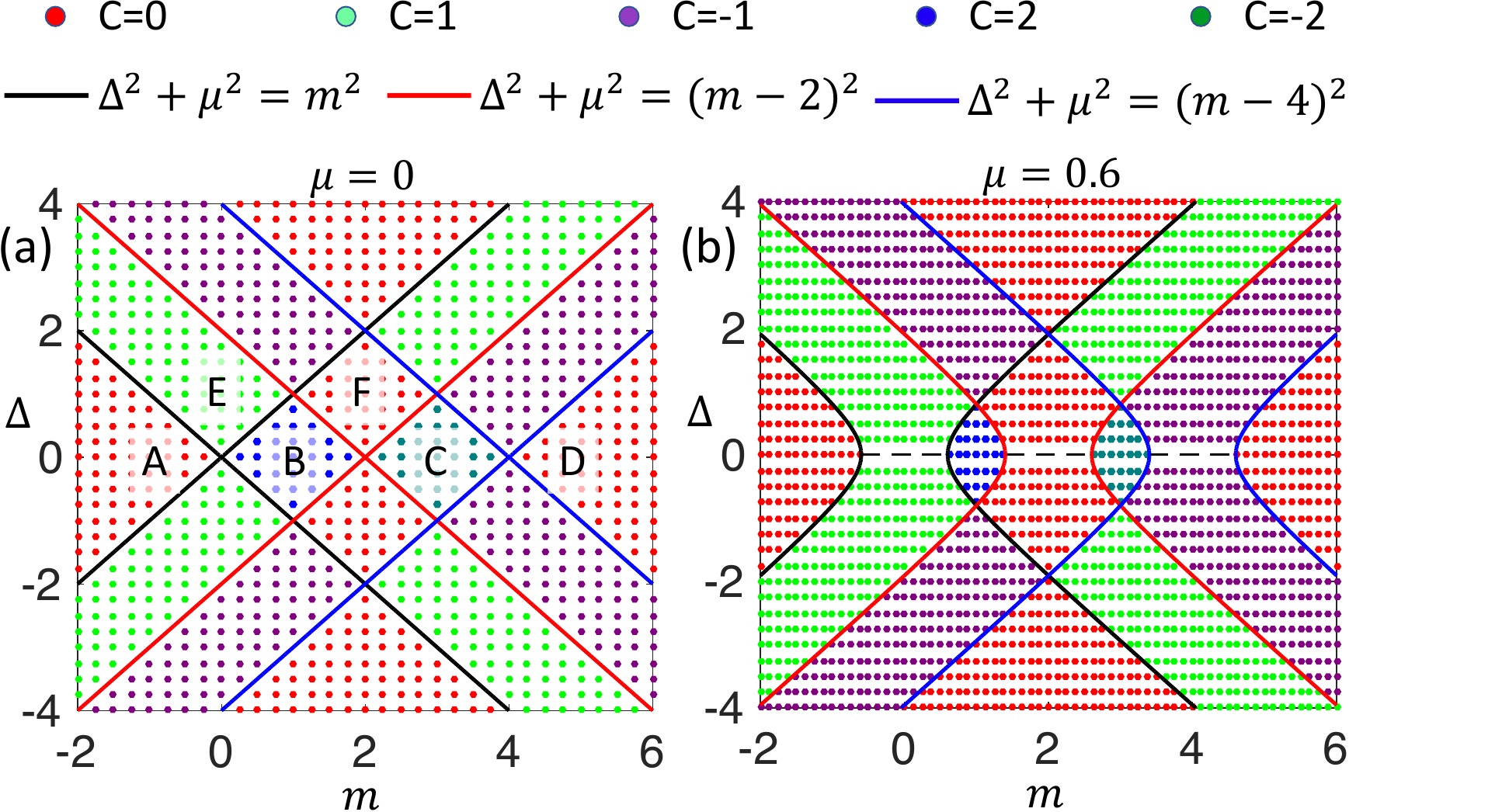}
	\caption{(a) The topological phase diagram of the SC Chern insulator Eq. (\ref{E3}) with $\mu=0$. As indicated in the legend, regions with different Chern numbers are filled with points with different colors. The expressions of three pairs of phase boundaries are also given in the legend ($\mu=0$ for the sub-figure (a)). (b) The topological phase diagram of the SC Chern insulator Eq. (\ref{E3}) with $\mu=0.6$. The sub-figure (b) shares the same legend with the sub-figure (a), but we have $\mu=0.6$ in the expressions of the phase boundaries for (b). The gray dashed lines within the regions $m\in(-\mu,\mu)$, $m\in (2-\mu,2+\mu)$, and $m\in (4-\mu,4+\mu)$ at $\Delta=0$ reflect the physics that the chemical potential is tuned to the conduction band, and zero paring strength leads to gapless spectra. Thus, we cannot define Chern number for points on these gray dashed lines.}
	\label{F2}
\end{figure} 
In the above section, we have demonstrated the close connections between the Chern insulator Eq. (\ref{E1}) and the Weyl semimetal Eq. (\ref{E2}). As a matter of fact, it turns out that we can obtain the VBSs of SC Weyl semimetals from the VBSs of SC Chern insulators. Because the VBSs of SC Chern insulators are determined by their topological properties, in this section we will delve into the topological phase diagram of SC Chern insulator Eq. (\ref{E3}), where  $H_{N}(\bm{k})$ refers to the Chern insulators described by Eq. (\ref{E1}). 

\subsection{The $\mu=0$ case}
For clarity, we first consider the $\mu=0$ case. The topological phase diagram of the continuous version of Eq. (\ref{E3}) is analyzed in a seminal paper about SC Chern insulators \cite{PhysRevB.82.184516}. Analogous to the arguments in this seminal paper, we will derive the topological phase diagram by checking the topological transitions at the gapless phase boundaries. First we give the BdG Chern numbers for the points on the $\Delta=0$ line. When $\Delta=0$, the BdG Hamiltonian Eq. (\ref{E3}) is decoupled into an upper left $2\times2$ block and a lower right $2\times2$ block. It is easy to find that insulators of the two blocks have the same Chern number. Thus, compared to the Chern number of the Chern insulators in Eq. (\ref{E1}), the BdG Chern number of SC Chern insulators at $\Delta=0$ simply doubles. As shown in Fig. \ref{F2}(a), we obtain $C=0$ for $m<0$ or $m>4$ (region A and region D), $C=2$ for $0<m<2$ (region B), and $C=-2$ for $2<m<4$ (region C).  

The Chern number of a SC Chern insulator doesn't change unless the bulk gap closes and reopens, so we determine the topological phase boundaries from gap closing conditions and derive the BdG Chern numbers in other regions by observing the changes of Chern number around the phase boundaries. The bulk spectra of the SC Chern insulator Eq. (\ref{E3}) are $E=\pm \sqrt{(\sin^{2} k_{x}+\sin^{2} k_{y})+(\Delta\pm m(\bm{k}))^{2}}$, in which we have used the notation $m(\bm{k})=m-(2-\cos k_{x}-\cos k_{y})$.
\begin{table}[h] \centering 
	\begin{tabular}{|c|c|c|} \hline $(k_{x}, k_{y})$ & $m(\bm{k})=m-(2-\cos k_{x}-\cos k_{y})$ & $\Delta^{2}$ \\ \hline $(0,0)$ & $m$ & $m^{2}$ \\ \hline $(0,\pi), (\pi, 0)$ & $m-2$ & $(m-2)^{2}$ \\ \hline $(\pi, \pi)$ & $m-4$ & $(m-4)^{2}$ \\ \hline \end{tabular} 
	\caption{Topological phase boundaries determined from the bulk gap closing condition (the $\mu=0$ case). When $\Delta=\pm m$, the gapless point is $(0,0)$. When $\Delta=\pm (m-2)$, the gapless points are $(0,\pi)$ and $(\pi,0)$. When $\Delta=\pm (m-4)$, the gapless point is $(\pi,\pi)$. } \label{T1} 
\end{table}
The bulk gap closing conditions require $\sin k_{x}=0$, $\sin k_{y}=0$, and $\Delta\pm m(\bm{k})=0$. According to the three equations, we obtain the topological phase boundaries, and the results are summarized in Table \ref{T1}.  When $\Delta=\pm m$, the gapless point is $(0,0)$. If $\Delta=\pm (m-2)$, the bulk gap closes at two points $(0,\pi)$ and $(\pi,0)$. For $\Delta=\pm (m-4)$, the bulk gap closes at $(\pi,\pi)$. Such three phase boundaries are three pairs of lines with different colors in Fig. \ref{F2}(a). For points not on the phase boundaries, the bulk spectra are fully gapped, and there is no topological phase transition. Now we derive the BdG Chern numbers in regions where $\Delta\neq0$. As shown in Fig. \ref{F2}(a), region E is adjacent to both region A and region B. The phase boundary between region E and region A is $\Delta=-m$, which corresponds to only one gapless point $(0,0)$, so the change of Chern number could be $1$ or $-1$. Because the Chern number in region A is $C=0$, we have $C=1$ or $C=-1$ in region E. On the other hand, the phase boundary between region E and region B is $\Delta=m$, corresponding to one gapless point $(0,0)$, so the change of Chern number is also $1$ or $-1$. We know that $C=2$ in region B, so in region E we have $C=1$ or $C=3$. Combining the two results, we conclude that the Chern number in region E is $C=1$. We obtain the Chern number in region F in a very similar way. The phase boundary between region B and region F is $\Delta=-(m-2)$, and bulk gap closes at two points $(0,\pi)$ and $(\pi, 0)$. We thus expect that the change of Chern number is $2$ or $-2$. Considering 
$C=2$ in region B, it could be $C=0$ or $C=4$ in region F. Given that the phase boundary between region F and region C also corresponds to two gapless points,
the change of Chern number is also $2$ or $-2$. 
Because $C=-2$ in region C, we have $C=0$ or $C=-4$ in region F.  Taking the two aspects into account, we conclude $C=0$ in region F. Similarly, we can analytically  obtain the Chern numbers in all regions of the topological phase diagram Fig. \ref{F2}(a).
\begin{table}[h] \centering 
	\begin{tabular}{|c|c|c|} \hline $(k_{x}, k_{y})$ & $m(\bm{k})=m-(2-\cos k_{x}-\cos k_{y})$ & $\Delta^{2}+\mu^{2}$ \\ \hline $(0,0)$ & $m$ & $m^{2}$ \\ \hline $(0,\pi), (\pi, 0)$ & $m-2$ & $(m-2)^{2}$ \\ \hline $(\pi, \pi)$ & $m-4$ & $(m-4)^{2}$ \\ \hline \end{tabular} 
	\caption{Topological phase boundaries determined from the bulk gap closing condition (the $\mu\neq0$ case). When $\Delta^{2}+\mu^{2}=m^{2}$, the gapless point is $(0,0)$. When $\Delta^{2}+\mu^{2}=(m-2)^{2}$, the gapless points are $(0,\pi)$ and $(\pi,0)$. When $\Delta^{2}+\mu^{2}=(m-4)^{2}$, the gapless point is $(\pi,\pi)$. } \label{T2} 
\end{table}

The analytical results above can be verified through numerical calculations.  We adopt the following formula to calculate the Chern number \cite{PhysRevLett.116.046401,PhysRevB.97.020501}. 
\begin{equation}
	\begin{aligned}
	C&= \int \frac{d\bm{k}}{2\pi} \rm{Im}\sum_{a,b} \frac{\langle u_{a}\vert\partial_{k_{x}}H \vert u_{b}\rangle\langle u_{b}\vert\partial_{k_{y}}H \vert u_{a}\rangle-(k_{x}\leftrightarrow k_{y})}{(E_{a}-E_{b})^{2}}.
	\label{E4}
	\end{aligned}
\end{equation}
For an insulator with $N$ bands in total and $M$ bands occupied, the summation of the index $a$ is from $1$ to $M$ (occupied bands), and the summation of the index $b$ is from $M+1$ to $N$ (the unoccupied bands). Applying this formula to our SC Chern insulator, we obtain the Chern numbers for the uniformly distributed points in the topological phase diagram Fig. \ref{F2}(a), where points with different Chern numbers are marked in different colors. Obviously, the numerical results are consistent with the analytical results, and points with different colors are separated by the analytical phase boundaries. 

\subsection{The $\mu\neq0$ case}
Then we calculate the topological phase diagram for the $\mu\neq0$ case. We still calculate the topological phase boundaries by applying the gap closing conditions. Since $\mu\neq0$, the bulk spectra now become 
\begin{widetext}
	\begin{equation}
		E=\pm \sqrt{(\sin^{2} k_{x}+ \sin^{2} k_{y})+\Delta^{2}+m(\bm{k})^{2}+\mu^{2}\pm2\sqrt{(\Delta^{2}+\mu^{2})m^{2}(\bm{k})+\mu^{2}(\sin^{2} k_{x}+ \sin^{2} k_{y})}},
	\end{equation}
\end{widetext}
where $m(\bm{k})=m-(2-\cos k_{x}-\cos k_{y})$. When we consider the case $m^{2}(\bm{k})-\mu^{2}\geq0$,  the gap closing conditions lead to $\sin^{2} k_{x}+\sin^{2}k_{y}+\Delta^{2}+\mu^{2}-m^{2}(\bm{k})=0$ and $(\sin^{2} k_{x}+\sin^{2}k_{y})(m^{2}(\bm{k})-\mu^{2})=0$. The two equations can be transformed to another three equations $\sin k_{x}=0$, $\sin k_{y}=0$, and  $\Delta^{2}+\mu^{2}-m^{2}(\bm{k})=0$. We obtain the topological phase boundaries from the equations above, and the results are given in Table \ref{T2}. There are three pairs of phase boundaries, and all of them are hyperbolas. 
\begin{figure*}[tbph]
	\centering
	\includegraphics[scale=0.407]{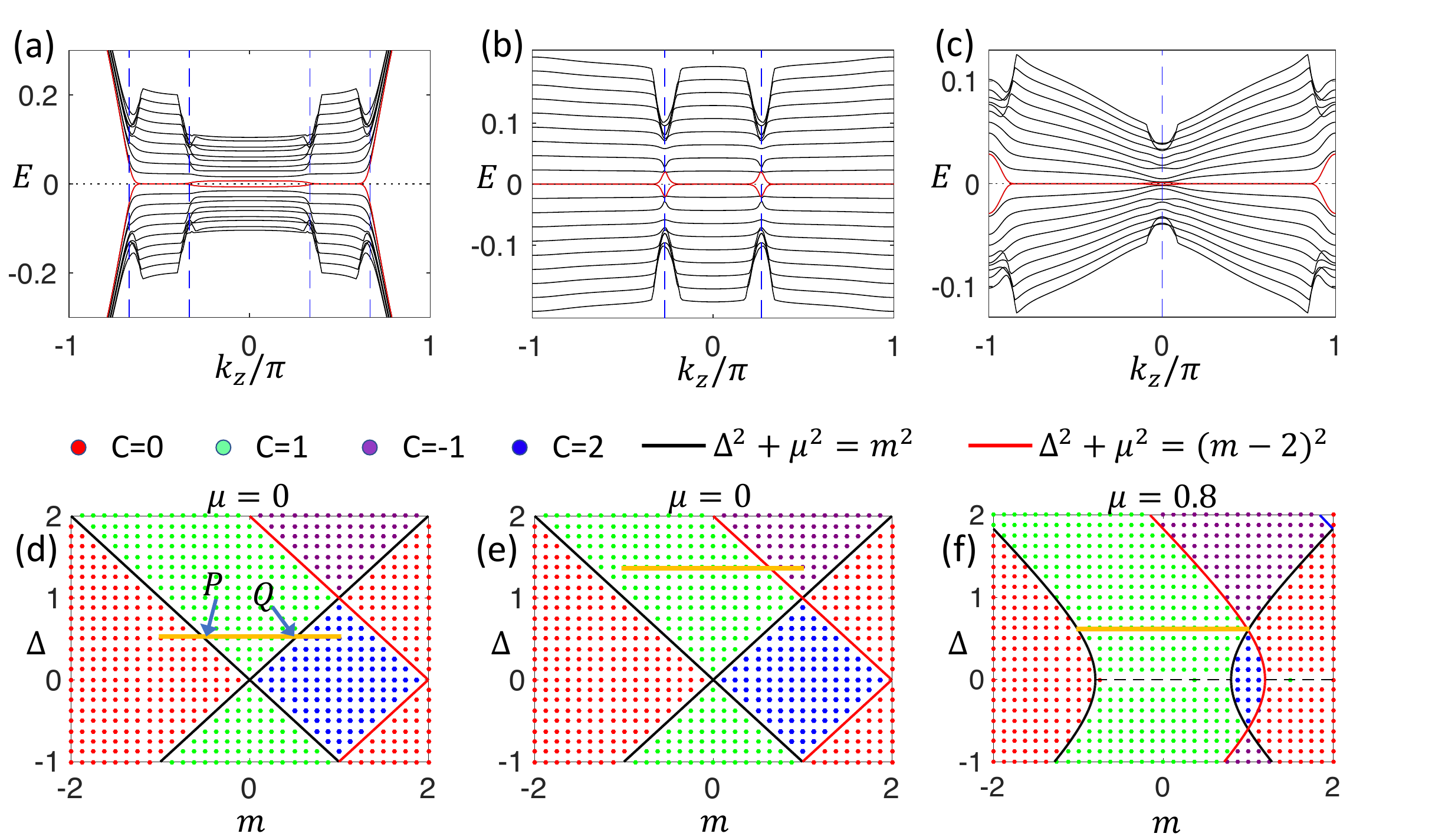}
	\caption{The VBSs of SC Weyl semimetal and the understanding from the topological phase diagram of corresponding SC Chern insulators. (a) The VBSs of SC Weyl semimetals with $\mu=0$ and $\Delta=0.5$. The vertical blue dashed lines are the analytical (exact) boundaries of the MFBs obtained from the corresponding phase diagram of SC Chern insulators in the sub-figure (d), and they are located at $k_{z}=\pm \pi/3$ and $k_{z}=\pm 2\pi/3$.  (b) The VBSs of SC Weyl semimetals with $\mu=0$ and $\Delta=4/3$. The vertical blue dashed lines are the analytical boundaries of the MFBs located at $k_{z}=\arccos 2/3\approx\pm0.27\pi.$ (c) The VBSs of SC Weyl semimetals with $\mu=0.8$ and $\Delta=0.6$. The analytical boundaries of the MFBs are located at $k_{z}=\pi$ and $k_{z}=0$ (the vertical blue dashed line). (d) The topological phase diagram of SC Chern insulator Eq. (\ref{E1}) with $\mu=0$. The thick orange line at $\Delta=0.5$ represents the SC Weyl semimetal in the sub-figure (a). Points $P$ and $Q$ explain the appearance of four bulk gapless points in Fig. \ref{F1}(c2). (e) The topological phase diagram of SC Chern insulator with $\mu=0$. The thick orange line at $\Delta=4/3$ represents the SC Weyl semimetal in the sub-figure (b). (f) The topological phase diagram of SC Chern insulator with $\mu=0.8$. The thick orange line at $\Delta=0.6$ represents the SC Weyl semimetal in the sub-figure (c).}
	\label{F3}
\end{figure*} 
As a concrete example, the topological phase diagram for the $\mu=0.6$ case is shown in Fig. \ref{F2}(b). Points with all kinds of colors calculated numerically with the formula Eq. (\ref{E4}) have different Chern numbers, which is demonstrated in the legend. Three pairs of hyperbolas with different colors are the aforementioned topological phase boundaries, and the explicit expressions of them are also given in the legend. We notice that the points with different Chern numbers are separated by the analytical topological phase boundaries, and numerical results are consistent with the analytical phase boundaries.   Compared to the topological phase diagram of the $\mu=0$ case, we find that for regions around $\Delta=0$, the deformed phase boundaries enlarge 
the regions with Chern number $C=1$ (region E in Fig. \ref{F2}(a)) and $C=-1$ and make regions with $C=2$ (region B) and $C=-2$ (region C) to shrink. In fact, if we further increase $\mu$, the regions with $C=2$ and $C=-2$ will get smaller. 

It is worthwhile to point out that the above topological phase boundaries are calculated with the assumption $m^{2}(\bm{k})-\mu^{2}\geq0$. Now we discuss the physics under the condition $m^{2}(\bm{k})-\mu^{2}<0$. This condition corresponds to three gray dashed lines within the regions $m\in(-\mu,\mu)$, $m\in (2-\mu,2+\mu)$, and $m\in (4-\mu,4+\mu)$ at $\Delta=0$ in Fig. \ref{F2}(b).  In fact, when the condition $m^{2}(\bm{k})-\mu^{2}<0$ is satisfied, the chemical potential is tuned to the conduction band, so zero paring strength ($\Delta=0$) leads to gapless spectra. Thus, we cannot define Chern number for points on this line. However, these gapless spectra don't correspond to topological phase transitions, because these gapless points are on the Fermi surface, and the dispersion around each point is not a Dirac cone. As shown in Fig. \ref{F2}(b), the two regions separated by a gray dashed line have the same Chern number. 

\section{Majorana Flat Bands in the Vortex Line of Superconducting Weyl Semimetals}
In this section we will present the VBSs of SC time reversal symmetry breaking Weyl semimetals Eq. (\ref{E3}), in which $H_{N}(\bm{k})$ refers to the normal state Weyl semimetal Eq. (\ref{E2}).  As shown in Fig. \ref{F3}, we have calculated the VBSs of SC Weyl semimetals under three sets of parameters, which  illustrates the universality of the appearance of MFBs. We also understand these MFBs and determine the boundaries of them from the  topological phase diagrams of the corresponding SC Chern insulators. 

\subsection{The $\mu=0$ and $\Delta=0.5$ Case}
To be specific, Fig. \ref{F3}(a) shows the VBSs of SC Weyl semimetals with $\mu=0$ and $\Delta=0.5$. The states close to the Fermi energy $E=0$ are colored red to make them noticeable.  We find that there are two pieces of zero energy flat bands ( the MFBs) enclosed by the blue dashed boundaries ($k_{z}=\pm\frac{2\pi}{3}$ and $k_{z}=\pm \frac{\pi}{3}$), which are obtained from the topological phase diagram of the corresponding SC Chern insulators in Fig. \ref{F3}(d). 

Now we discuss how to understand the appearance of MFBs and obtain the exact boundaries of them from the topological phase diagram Fig. \ref{F3}(d). Because there is a direct correspondence between the Chern insulator Eq. (\ref{E1}) and the Weyl semimetal Eq. (\ref{E2}) by identifying $m$ in the Chern insulator with $2t_{z}\cos k_{z}$ in the Weyl semimetal, a fixed $k_{z}$ in the Weyl semimetal gives a Chern insulator with $m=2t_{z}\cos k_{z}$, so the SC Weyl semimetal can be described by the thick orange line in the topological phase diagram Fig. \ref{F3}(d). We have set $t_{z}=0.5$, so $2t_{z}\cos k_{z}\in [-1,1]$, which is the range of $m$ of the orange line. On the other hand, since we are studying the VBSs of a SC Weyl semimetal with $\Delta=0.5$, the $y$-coordinate of the orange line is $\Delta=0.5$. The orange line intersects the topological phase boundaries $\Delta^{2}=m^{2}$ at $(0.5,0.5)$ and $(-0.5,0.5)$. Thus, the $C=1$ part (the green region) of the orange line is within the range $m\in (-0.5,0.5)$. Then the condition $\cos k_{z}\in (-0.5, 0.5)$ gives rise to $k_{z}\in (-2\pi/3,-\pi/3)$ and $k_{z}\in(\pi/3, 2\pi/3)$. Because the chiral superconductor with odd Chern number has a MZM in the vortex core \cite{PhysRevB.61.10267,PhysRevB.82.184516},  we conclude that for $k_{z}\in (-2\pi/3,-\pi/3)$ and $k_{z}\in(\pi/3, 2\pi/3)$,  the MZMs in the VBSs at the two ranges of $k_{z}$ form the MFBs. Obviously, the exact boundaries of the MFBs are $k_{z}=\pm \pi/3$ and $k_{z}=\pm 2\pi/3$, which are the blue dashed boundaries in Fig. \ref{F3}(a). We would like to point out that the gap closing condition $m=\pm \Delta$ in the topological phase diagram corresponds to the bulk gap closing condition $\cos k_{z}=\pm \Delta$ of the SC Weyl semimetal mentioned in Fig. \ref{F1}(c2). The phase transition point $P$ ($m=-\Delta$) and $Q$ ($m=\Delta$) in  Fig. \ref{F3}(d) leads to the bulk gap closing points $P_{1,2}$ (solutions of $\cos k_{z}=-\Delta$) and $Q_{1,2}$ (solutions of $\cos k_{z}=\Delta$) in Fig. \ref{F1}(c2).  

\subsection{The Hybridization between Edge MZM and Vortex Core MZM}
Above paragraph gives the exact boundaries of MFBs, but we can find in Fig. \ref{F3}(a) that there is a noticeable discrepancy between the exact boundaries (the blue dashed lines) and the region of MFBs, and the MFBs fail to reach the exact phase boundaries. 
\begin{figure}[tbph]
	\centering
	\includegraphics[scale=0.412]{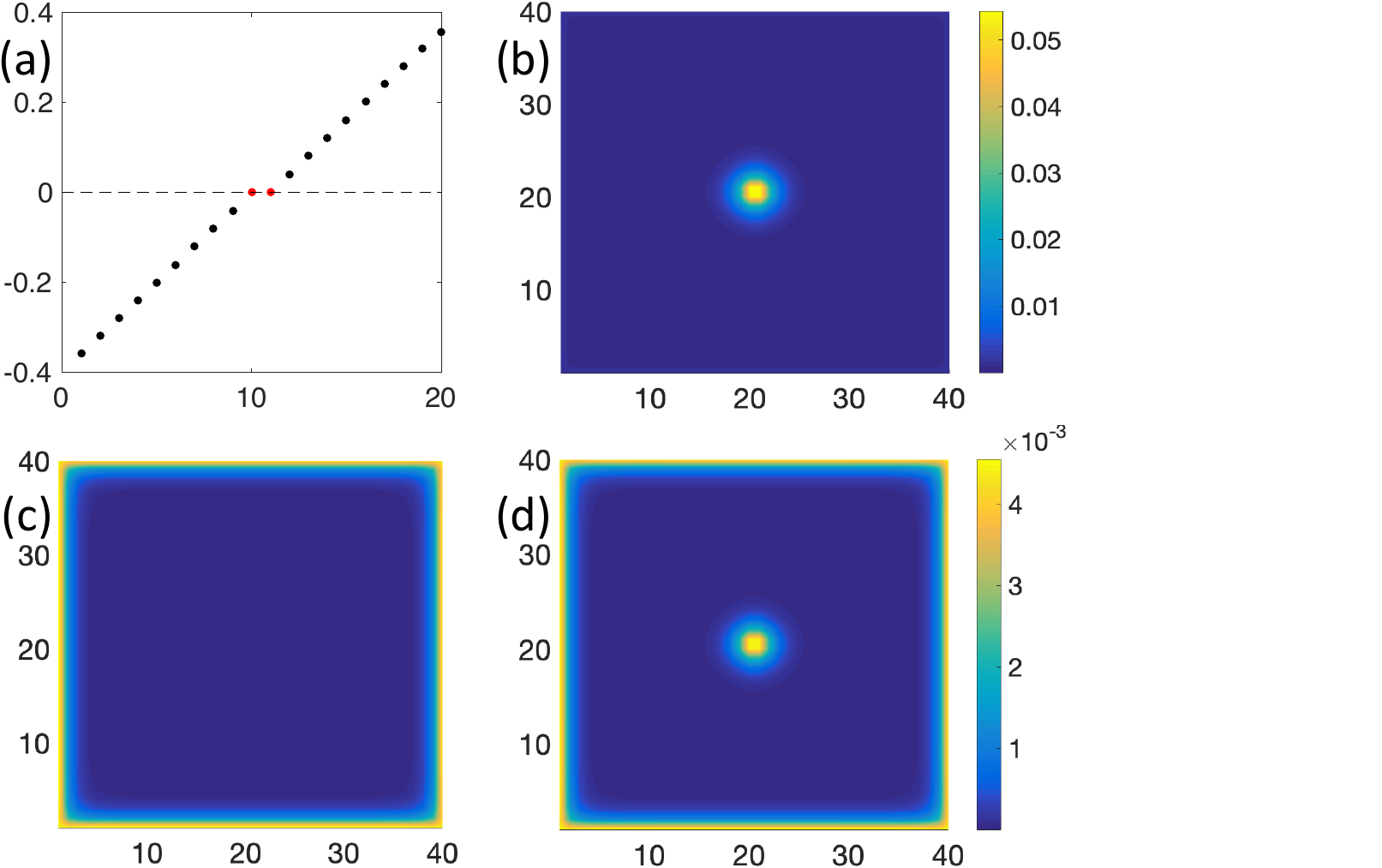}
	\caption{(a) is the spectra of VBSs of SC Weyl semimetal at $k_{z}=0.5\pi$, and other parameters are set as $\mu=0$, $\Delta=0.5$, lattice size $N=40$. The two degenerate zero-energy states are colored in red. We can construct the wave function localized at the vortex core (sub-figure (b)) and the wave function localized at the edge (sub-figure (c)) through the linear superposition of the eigen-functions of the two zero-energy states. A typical eigen-function of the zero-energy state is shown in (d), and the wave function appears at both the vortex core and the edge.}
	\label{F3.5}
\end{figure} 
Besides, we know that there is \textit{one} MZM in the vortex core of a chiral topological
superconductor, but there are \textit{two} degenerate MZMs in the VBSs of each $k_{z}$, which is shown in Fig. \ref{F3}(a). To solve the puzzles, the key point is to realize that there is one MZM localized at the vortex core  and one MZM localized at the edge in the VBSs at each $k_{z}$, and the hybridization between the edge MZM and vortex core MZM breaks  the degeneracy of the two MZMs. 
For concreteness, we fix $k_{z}=0.5\pi$ in Fig. \ref{F3}(a), and the spectra consist of discrete energy levels, which is shown in Fig. \ref{F3.5}(a). Obviously, there are two degenerate MZMs (colored in red). We claim that one of the two MZMs is localized at the vortex core, while the other is localized at the edge, and what follows is a brief argument. Since the BdG Chern number of a two-dimensional model obtained at $k_{z}$ with  $k_{z}\in (\pi/3,2\pi/3)$ is $C=1$, the open boundary calculations  give edge states with energy levels $\epsilon_{n}=(n+1/2)\epsilon_{0}$, where $n=0,\pm1,\pm2,\cdots$ \cite{Alicea_2012}. After we insert a $\pi$-flux, the energy levels will shift by half an integer, and the new spectra become $\epsilon_{n}=n\epsilon_{0}$ \cite{Alicea_2012}, containing the zero-energy state. Thus, an edge MZM will emerge after the $\pi$-flux is inserted (see Fig. \ref{F3.5}(c)). On the other hand, since the BdG Chern number is $C=1$ for the two-dimensional model at $k_{z}=0.5\pi$, there is also one MZM localized at the vortex core (see Fig. \ref{F3.5}(b)). So we conclude that there are doubly degenerate MZMs in the VBSs.   

However, when we plot the numerically calculated eigen-functions $\vert\psi_{1}\rangle$ and $\vert \psi_{2}\rangle$ corresponding to the two zero-energy states in Fig. \ref{F3.5}(a), the wavefunction appears at both the edge and the vortex core (like Fig. \ref{F3.5}(d)), which is different from the wavefuntion of an edge MZM or a vortex core MZM. The point is that the doubly degenerate zero-energy states form a two-dimensional space, and $\vert\psi_{1}\rangle$ and $\vert\psi_{2}\rangle$ are the orthogonal normalized basis vectors. We denote the wavefunctions of edge MZM and vortex core MZM as $\vert \psi_{e}\rangle$ and $\vert \psi_{c}\rangle$, respectively.  $\vert \psi_{e}\rangle$ and $\vert\psi_{c}\rangle$ constitute another set of orthogonal normalized basis vectors. Because the two sets of basis vectors $\{\vert\psi_{1}\rangle,\vert\psi_{2}\rangle\}$ and $\{\vert\psi_{e}\rangle,\vert\psi_{c}\rangle\}$ span the same two-dimensional space, we expect a $U(2)$ matrix to connect them. Indeed we can find such a $U(2)$ matrix. We linearly superpose $\vert\psi_{1}\rangle$ and $\vert\psi_{2}\rangle$ to obtain another set of basis vectors, which is represented by the following equation
\begin{equation}
\begin{pmatrix}
    \vert\psi_{e}\rangle\\
\vert\psi_{c}\rangle
\end{pmatrix}=\frac{1}{\sqrt{2}}\begin{pmatrix}
    1& e^{i\theta}\\
    1& -e^{i\theta}
\end{pmatrix}\begin{pmatrix}
    \vert\psi_{1}\rangle\\
\vert\psi_{2}\rangle
\end{pmatrix}.
\end{equation}
It is easy to check that the $2\times2$ matrix in the above equation is a unitary matrix. We find that there always exists a $\theta\in[0, 2\pi)$ such that $\vert\psi_{e}\rangle$ is localized at the edge and $\vert\psi_{c}\rangle$ is localized at the vortex core. Thus, we conclude that there are an edge MZM and a vortex core MZM in the VBSs. Because the two MZMs are degenerate, the orthogonal normalized basis vectors calculated by the numerical method ($\vert\psi_{1}\rangle$ and $\vert\psi_{2}\rangle$) could be the linear superposition of the edge MZM and the vortex core MZM, and the wavefunctions appear at both the edge and the flux core (see Fig. \ref{F3.5}(d)). 

Because the edge MZM is localized at the edge, and the vortex core is localized at the vortex core, when the lattice size is large enough, there is almost no overlap between the wavefunction of the edge MZM and the vortex core MZM. In this case, we may assume that the two MZMs are degenerate.  However, things are different when $k_{z}$ is near the boundaries ($k_{z}=\pm \pi/3$ or $k_{z}=\pm 2\pi/3$) of MFBs in Fig.  \ref{F3}(a). As we tune $k_{z}$ to the boundary, the corresponding SC Chern insulator approaches  the phase transition point, and the widths of the edge states increase. At the phase transition point, the widths of the edge states diverge \cite{doi:10.1143/JPSJ.77.031007,PhysRevB.82.184516}. Thus, when we tune $k_{z}$ to the boundaries of the MFBs, the edge MZMs are not well localized.  As a result, there is finite overlap between the edge MZM and the vortex core MZM, giving rise to the splitting of the doubly degenerate MZMs. This is the reason why the MFBs fail to reach the exact phase boundaries and there is a finite gap between two low-energy states (marked in red) near the exact phase boundaries. This physical picture can be verified by observing the behavior of MFBs under change of lattice size.   Fig. \ref{AF2}(b) and (d) in the Appendix show that for a given $k_{z}$ near the phase boundary, as we increase the lattice size, the gap between the two low-energy states decrease exponentially. When we increase the lattice size $N$, the distance between the vortex core and the edge gets larger, and there is less overlap between the wavefunctions of edge MZM and vortex core MZM, naturally resulting in a smaller splitting of the two degenerate MZMs. Fig. \ref{AF2}(a) and (c) show that as we increase the lattice size, the gap between two low-energy states around the phase boundaries gets smaller, and the MFBs are closer to the exact phase boundaries.

\subsection{The Effect of $\Delta$ and $\mu$ on the MFBs}
We have discussed the properties of MFBs for given values of $\Delta$ and $\mu$. In this subsection, we will study the effect  of parameters $\Delta$ and $\mu$ on the MFBs.  First, we keep $\mu=0$ and change $\Delta$ to be $4/3$, and the corresponding VBSs are shown in Fig.\ref{F3}(b). Such a SC Weyl semimetal is described by the thick orange line in Fig. \ref{F3}(e). In this case, the Chern numbers for SC  insulators obtained by fixing $k_{z}$ in the SC Weyl semimetals are either $C=1$ or $C=-1$. However, the orange line intersects the topological phase boundary $\Delta=-(m-2)$ at ($2/3,4/3$), where the spectrum of the SC Chern insulator is gapless, and we cannot define the BdG Chern number. $m=2/3$ corresponds to $2t_{z}\cos k_{z}=2/3$, and we obtain $k_{z}=\arccos 2/3\approx \pm 0.27\pi$. Thus, for SC insulators obtained at $k_{z}\neq 0.27\pi$ in the SC Weyl semimetal, the BdG Chern numbers are odd, and vortex MZMs and the edge MZMs make up the MFB. Because the hybridization between the edge MZM and the vortex MZM is significant around the phase boundaries, we find finite gaps between the two low-energy states in the VBSs near the phase boundaries.  

The MFB on the whole $k_{z}$ axis in Fig. \ref{F3}(b) is obtained by tuning the pairing strength $\Delta$, and we can also obtain the MFB on the whole $k_{z}$ axis by changing the chemical potential $\mu$. As shown in Fig. \ref{F3}(c),  there is MFB on the whole $k_{z}$ axis in the VBSs of SC Weyl semimetal with $\mu=0.8$ and $\Delta=0.6$, which is represented by the thick orange line in Fig. \ref{F3}(f). The orange line intersects the topological phase boundaries $\Delta^{2}+\mu^{2}=m^{2}$ at $(-1,0.6)$ and $(1,0.6)$. The intersections at $m=1$ and $m=-1$ corresponds to $k_{z}=0$ and $k_{z}=\pm\pi$, respectively. Thus, if $k_{z}\neq0$ or $k_{z}\neq\pm\pi$, the corresponding two-dimensional SC Chern insulators have Chern number $C=1$, and the vortex core MZMs and the edge MZMs form the MFB on the $k_{z}$ axis.  $k_{z}=0$ and $k_{z}=\pm\pi$ are the phase transition points of SC Chern insulators. When $k_{z}$ is near these points, the hybridization of the vortex core MZM and the edge MZM also leads to a noticeable gap between the two low-energy states.  

\section{Realizing Superconducting Weyl Semimetals by Considering Attractive Interactions}
Above discussions are based on the BdG Hamiltonian Eq. (\ref{E3}), in which the pairing term is added by hand. A natural question arises that how can we obtain such a paring term. In the following texts, we will show that under appropriate parameters, the mean field approach can give the BCS pairing from the model of Weyl semimetal considering the attractive Hubbard interaction. 
\begin{equation}
	\begin{aligned}
		H_{\rm{Hubbard}}&=\sum_{\bm{k}} \psi^{\dagger}(\bm{k}) (H_{\rm{WSM}}(\bm{k})-\mu) \psi (\bm{k})\\
		&-\frac{U}{V}\sum_{\bm{q}}\sum_{\bm{k}}\sum_{\bm{k'}} c^{\dagger}_{\bm{k'}+\bm{q},\uparrow}c^{\dagger}_{-\bm{k'},\downarrow}c_{-\bm{k},\downarrow}c_{\bm{k}+\bm{q},\uparrow}.
		\label{E6}
	\end{aligned}
\end{equation}
\begin{figure*}[tbph]
	\centering
	\includegraphics[scale=0.607]{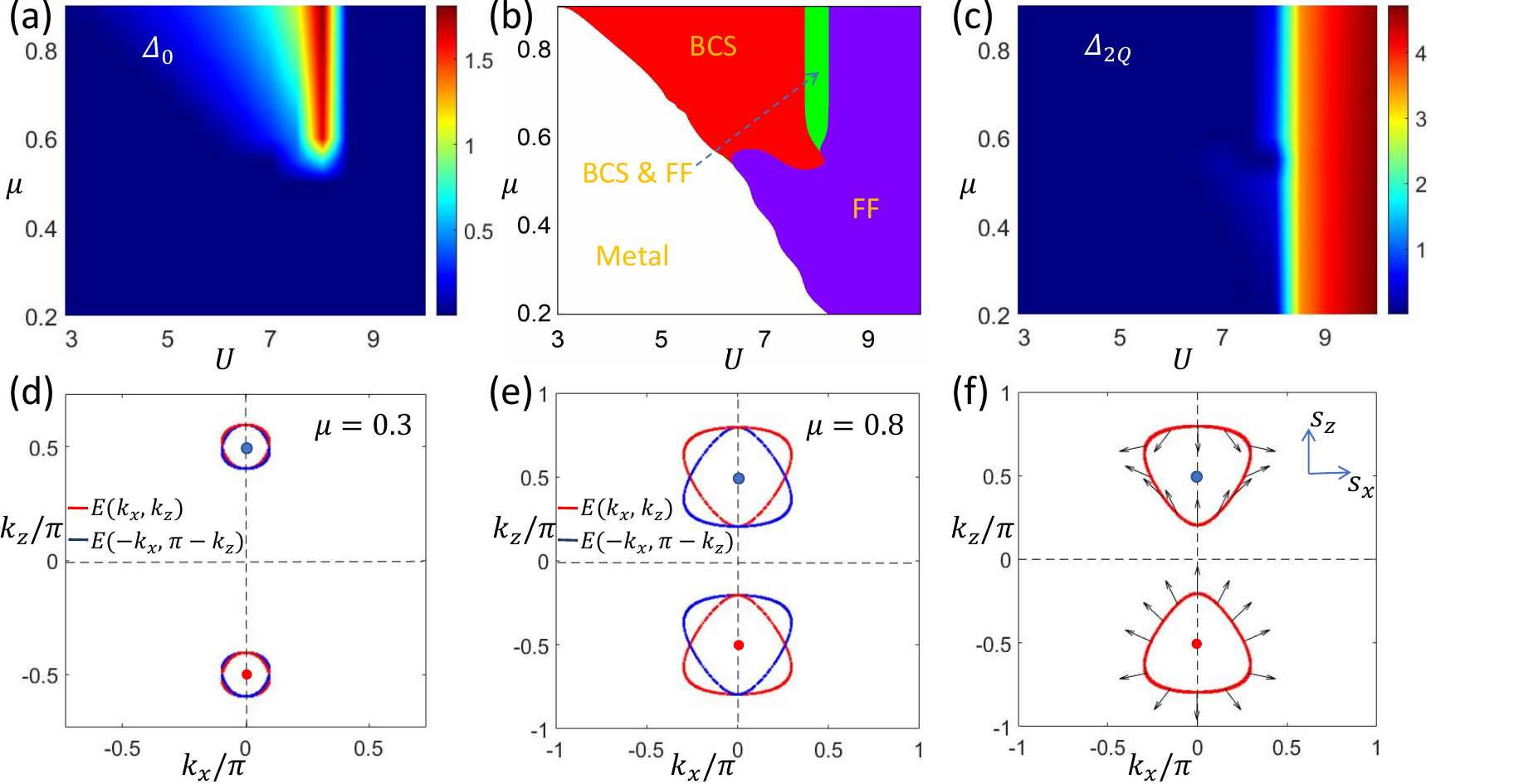}
	\caption{The sub-figures (a) and (c) are the distributions of order parameters $\Delta_{\bm{0}}$ and $\Delta_{\bm{2Q}}$ in the $U-\mu$ space, respectively. Combining (a) and (c), we can obtain the phase diagram (b) of the model of Weyl semimetal with attractive Hubbard interaction. The sub-figures (d) and (e) show the Fermi surfaces and their particle-hole counterparts of the normal state Weyl semimetal at $\mu=0.3$ and $\mu=0.8$ (with $k_{y}=0$), respectively. (f) shows the Fermi surface of the Weyl semimetal at $\mu=0.8$ and the spin textures on it ($k_{y}=0$).}
	\label{F4}
\end{figure*} 
As shown in Fig. \ref{F4}(f), the Fermi surfaces of the normal state Weyl semimetal consist of two closed circles around two Weyl points. Thus, the pairing with momentum $\bm{q}=\bm{0}$ and $\bm{q}=2\bm{Q}$ may dominate,  where $\bm{Q}=(0,0,\pi/2)$ is the location of the Weyl point. $\bm{q}=\bm{0}$ corresponds to the BCS pairing between two Weyl cones, while $\bm{q}=2\bm{Q}$ corresponds to the Fulde-Ferrell (FF) pairing within each Weyl cone \cite{PhysRev.135.A550}.  So we keep only $\bm{q}=\bm{0}$ and $\bm{q}=2\bm{Q}$ terms in the summation of $\bm{q}$ in Eq. (\ref{E6}). 

To apply the mean field approximation, we choose the order parameters as  $\Delta_{\bm{q}}=-\frac{U}{V}\sum_{\bm{k}}\langle c_{-\bm{k},\downarrow}c_{\bm{k}+\bm{q},\uparrow}\rangle$, where $\bm{q}$ could be $\bm{0}$ or $2\bm{Q}$. When $\Delta_{0}$ coexists with $\Delta_{\bm{2Q}}$, we cannot write the model in the form of a BdG Hamiltonian under the basis $\Psi^{\dagger}(\bm{k})=(c^{\dagger}_{\bm{k}, \uparrow}, c^{\dagger}_{\bm{k}, \downarrow},c_{-\bm{k}, \uparrow}, c_{-\bm{k}, \downarrow})$ with $\bm{k}$ in the full Brillouin zone. Because the wave-vector $2\bm{Q}$ in the FF pairing $\Delta_{\bm{2Q}}$ is along $z$ direction, we consider the Brillouin zone folded in $z$ direction with $k_{x}, k_{y}\in (-\pi,\pi]$ and $k_{z}\in (0,\pi]$. The new basis considering the folded Brillouin zone (FBZ) is $\Psi^{\dagger}(\bm{k})=(\tilde{c}^{\dagger}_{\bm{k}},\tilde{c}^{\dagger}_{\bm{k}+2\bm{Q}},\tilde{c}_{-\bm{k}},\tilde{c}_{-\bm{k}-2\bm{Q}})$, where we have defined $\tilde{c}^{\dagger}_{\bm{k}}=(c^{\dagger}_{\bm{k},\uparrow}, c^{\dagger}_{\bm{k},\downarrow})$. Under the new basis, the Hamiltonian now takes the following form
\begin{equation}
	\begin{aligned}
		H&=\frac{V}{U}\vert{\Delta_{0}}\vert^{2}+\frac{V}{U}\vert{\Delta_{2\bm{Q}}}\vert^{2}+\frac{1}{2}\sum_{\bm{k}} \text{Tr} H_{N}(\bm{k})\\
		&+\frac{1}{2}\sum_{\bm{k}\in\text{FBZ}}\Psi^{\dagger}(\bm{k}) H_{\text{BdG}}(\bm{k})\Psi(\bm{k}),
	\end{aligned}
\end{equation} in which the explicit form of $H_{\text{BdG}}(\bm{k})$ is given in Eq. (\ref{EA10}) in Appendix A.  After writing down the BdG Hamiltonian, according to the definition of the order parameters  $\Delta_{\bm{0}}$ and $\Delta_{\bm{2Q}}$, we can derive the self-consistent gap equations, which are Eq. (\ref{EA13}) and Eq. (\ref{EA14}) in Appendix A. 

Then the phase diagram in the $\mu-U$ space can be obtained by solving the self-consistent gap equations under various parameters, and the results are shown in Fig. \ref{F4}(a)-(c). Fig. \ref{F4}(a) and (c) show the magnitude of $\Delta_{\bm{0}}$ and $\Delta_{\bm{2Q}}$ in the $\mu-U$ space, respectively. Combining the results in Fig. \ref{F4}(a) and (c), we can obtain the phase diagram Fig. \ref{F4}(b). In the BCS phase (the red region), $\Delta_{\bm{0}}$ dominates, and we have $\vert \Delta_{\bm{0}}\vert\gg\vert \Delta_{2\bm{Q}}\vert$ and $\vert \Delta_{2\bm{Q}}\vert<1\times 10^{-4}$. In contrast,  $\Delta_{2\bm{Q}}$ dominates in the FF phase (the purple region), and $\vert \Delta_{\bm{0}}\vert\ll \vert \Delta_{2\bm{Q}}\vert$ and $\vert \Delta_{\bm{0}}\vert<1\times 10^{-4}$. The BCS phase and FF phase coexist in the green region, where $\vert \Delta_{\bm{0}}\vert>1\times 10^{-4}$ and $\vert \Delta_{2\bm{Q}}\vert>1\times 10^{-4}$. When $\mu$ and $U$ are small (the white region), both the magnitude of $\Delta_{\bm{0}}$ and $\Delta_{\bm{2Q}}$ are negligibly small ($<1\times 10^{-4}$), and it is the normal-state metal phase. We notice that as we increase $\mu$, we need a smaller $U$ to obtain a SC phase. Since the low-energy  normal state physics is described by a 3D Weyl cone, straightforward calculations give the  density of states (DOS) $g(E)\propto E^{2}$. Thus, when we increase $\mu$, the DOS at the Fermi surface gets larger, which provides more electrons for forming Cooper pairs, so we need a smaller $U$ to arrive at the SC phase. 

There are still other remarkable features in the phase diagram Fig. \ref{F4}(b). First, the FF pairing is favored for smaller $\mu$, while the BCS pairing is favored for greater $\mu$. To understand this, we compare the Fermi surfaces in Fig. \ref{F4}(d) with $\mu=0.3$ and Fig. \ref{F4}(e) with $\mu=0.8$. In the two sub-figures, the blue closed lines are Fermi surfaces $E(k_{x},k_{y},k_{z})$, and the red lines are their particle-hole counterparts $E(-k_{x},-k_{y},\pi-k_{z})$ considering the paring momentum $2\bm{Q}$. As shown in Fig. \ref{F4}(e), for a greater $\mu$, there could be significant difference between $E(k_{x},k_{y},k_{z})$ and $-E(k_{x},-k_{y},\pi-k_{z})$, which  hinders the pairing with momentum $2\bm{Q}$.  On the other hand, because there is inversion symmetry in this model, we have $E(k_{x},k_{y},k_{z})=E(-k_{x},-k_{y},-k_{z})$. So the Fermi surfaces and their particle-hole counterparts considering zero momentum pairing always overlap, and the pairing with momentum $\bm{0}$ will not be weakened.    Thus, increasing the chemical potential  suppresses the FF phase with paring momentum $2\bm{Q}$, and the BCS pairing with momentum $\bm{0}$  dominates, corresponding to the BCS phase (the red region in Fig. \ref{F4}(b)). Second, we find that if $U$ is large enough, the FF pairing always dominates. Let's explain this by using the information given in Fig. \ref{F4}(f), where the spin textures are plotted on the Fermi surfaces. Because we are considering the singlet pairing, the pairing between electrons with opposite spins is favored. As shown in Fig. \ref{F4}(f), the spin of the electron at $(k_{x},k_{y},k_{z})$ is almost opposite to that of the electron at $(-k_{x},-k_{y},\pi-k_{z})$ for every point on the Fermi surfaces. However, only a small portion of the electrons on the Fermi surfaces satisfy the condition that the spin of the electron at $(k_{x},k_{y},k_{z})$ is opposite to that of the electron at $(-k_{x},-k_{y},-k_{z})$. As a result, the FF pairing between two electrons with total momentum $2\bm{Q}$ is favored. Although in the BCS phase (the red region) the FF pairing is suppressed due to the off-resonance between particle and hole bands, if $U$ is large enough, the interaction will compensate for such energy difference. In this case, the role of spin textures become dominant, and FF pairing is favored, which corresponds to the FF phase (the purple region) in Fig. \ref{F4}(b).   

The phase diagram Fig. \ref{F4}(b) is obtained by assuming the order parameters $\Delta_{\bm{0}}$ and $\Delta_{\bm{2Q}}$ are real.  Seriously speaking, such assumption is too strong. In fact, there could be phase differences between the two order parameters. However, if we solve the self-consistent gap equations with a phase difference between the two order parameters, we find that the main features of the phase diagram in the above paragraphs are preserved. For example, the self-consistent solutions for the parameters $\mu=0.8$ and $U=7$ in the phase diagram Fig. \ref{F4}(b) are $\Delta_{\bm{0}}=0.592$ and $\Delta_{\bm{2Q}}=0$. After considering the phase difference between the two parameters, the self-consistent solutions are still $\vert\Delta_{\bm{0}}\vert=0.592$ and $\vert\Delta_{\bm{2Q}}\vert=0$. We also find that for the points around $\mu=0.8$ and $U=7$, we still obtain BCS phase after considering the phase difference between $\Delta_{\bm{0}}$ and $\Delta_{\bm{2Q}}$. Thus, the BCS phase in the phase diagram Fig. \ref{F4}(b) are still reliable even if we consider the phase difference between the two parameters. 

\section{The $k_{z}$-dependent Chern-Simons Term}
In this section we will propose the topological invariants characterizing the aforementioned MFBs in the vortex line of SC Weyl semimetals. Because the vortex line along $z$ direction preserves the translation symmetry in $z$ direction, $k_{z}$ is a good quantum number. For this reason, we decompose the VBSs of SC Weyl semimetals into the VBSs of SC Chern insulators labelled by $k_{z}$. Similarly, the topological invariants characterizing the VBSs are also defined for each $k_{z}$. Since we care about the VBSs of SC Weyl semimetals, we will propose the topologial invariant to characterize the vortex core MZM in the MFBs.

When $k_{z}$ is fixed, we obtain a 2D SC Chern insulator belonging to class D, and the vortex line becomes a point defect.  According to the classification of topological defects \cite{PhysRevB.82.115120}, for a point defect in 2D, the Hamiltonian depends on two momentum variables and one position variable. In this case, the parameters in $\bm{k}$ space $k_{x}$, $k_{y}$, and the real space parameter $\phi$ make up the three-dimensional synthetic space $T^{2}\times S^{1}$, where $\phi$ is the polar angle in the real 2D space. It is shown that a point defect in class D is characterized by a $Z_{2}$ invariant determining  the presence or absence of Majorana zero modes \cite{PhysRevB.82.115120,PhysRevLett.119.047001}. The $Z_{2}$ invariant is the so-called Chern-Simons invariant 
\begin{equation}
	\nu=\frac{2}{2!}(\frac{i}{2\pi})^{2}\int_{T^{2}\times S^{1}}\mathcal{Q}_{3}\quad \text{mod}\quad 2.
\end{equation}
Here $\mathcal{Q}_{3}$ is the Chern-Simon form 
\begin{equation}
	\mathcal{Q}_{3}=\text{Tr}[\mathcal{A}d\mathcal{A}+\frac{2}{3}\mathcal{A}^{3}], 
\end{equation}
where $\mathcal{A}$ is the Berry's connection.

For a 2D SC Chern insulator (obtained by fixing $k_{z}$ in  SC Weyl semimetal Eq. (\ref{E3})) with BdG Chern number $C(k_{z})$ and a vortex with vorticity $n$, it is shown that the $Z_{2}$ invariant characterizing the vortex mode is \cite{PhysRevB.82.115120,PhysRevLett.119.047001}
\begin{equation}
	v(k_{z})=C(k_{z})n \quad \text{mod} \quad 2. 
	\label{E10}
\end{equation}
Throughout this article, we consider the vortex with $n=1$, thus Eq. (\ref{E10}) reduces to $\nu(k_{z})=C(k_{z})$ $\text{mod}$  $2$. That is, if the BdG Chern number $C(k_{z})$ of the SC Chern insulator obtained at $k_{z}$ in the SC Weyl semimetal is odd, there will be a Majorana zero mode in the vortex core, which is consistent with the arguments we use to explain the appearance of MFBs in the previous texts. Thus, the $k_{z}$-dependent Chern-Simon invariant Eq. (\ref{E10}) is the expected topological invariant for characterizing the MFBs in the vortex line of SC Weyl semimetals. 

\section{Summary and Discussions}
In summary, we have calculated the VBSs of SC Weyl semimetals which breaks time reversal symmetry and obtain the MFBs in the vortex line. To understand the appearance of the MFBs, we first show that the Weyl semimetal can be decomposed into stacked Chern insulators with different Chern numbers. Then we calculate the topological phase diagram of the corresponding SC Chern insulators through both analytical and numerical methods. By mapping the SC Weyl semimetal into a series of SC Chern insulators, we understand the MFBs in the vortex line of SC Weyl semimetal as the MZMs in the vortex line of SC Chern insulators with odd BdG Chern numbers. We show that the regions of MFBs are not consistent with the regions enclosed by the exact phase boundaries obtained from the topological phase diagram, and we attribute such inconsistency to the hybridization between the vortex-core MZM and the edge MZM at $k_{z}$ around the phase transition points. To realize the SC Weyl semimetal with BCS pairing, we consider the Weyl semimetal with attractive Hubbard interaction. We find that under appropriate parameters, the mean field calculations give rise to  the expected BCS pairing phase.  Finally, based on the previous study about the classification of topological defects, we propose the $k_{z}$-dependent $Z_{2}$ Chern-Simons invariant to characterize the MFBs.

It is worthwhile to point out that different from Weyl semimetals breaking time reversal symmetry, the VBSs of time reversal symmetric Weyl semimetals can host propagating gapless Majorana modes \cite{PhysRevLett.130.156402}. This is because the low-energy effective BdG Hamiltonian of a SC time reversal symmetric Weyl semimetal consists of mutually anti-commuting matrices. As a result, the spectra are fully gapped, and the in-gap gapless Majorana modes are protected by the emergent second Chern number. In our case, things are quite different.  The spectra of the effective BdG Hamiltonian of a SC time reversal symmetry breaking Weyl semimetal are gapless, and we cannot define the second Chern number to protect propagating gapless Majorana modes. Instead, we can study the VBSs for each $k_{z}$, and MZMs in a certain range of $k_{z}$ constitute the MFBs.  

\section*{Acknowledgments}
We appreciate the valuable discussions with Ting Fung Jeffrey Poon and Wei Jia about the calculations of Chern numbers and the mean-field method considering the folded Brillouin zone.  We also thank X.X. Wu, X.J. Liu, S.S. Qin, and J.S. Hong for helpful discussions.  

 \textit{Note added}: We recently noticed an independent work discussing the Majorana flat bands in the vortex line \cite{zhang2025doublemajoranavortexflat}. Different from our work, they consider the SC Dirac semimetals with unconventional pairing, and the corresponding double Majorana flat bands are protected by rotational symmetry. 

\appendix
\section{The Mean Field Approach}
The pairing term in the BdG Hamiltonian Eq. (\ref{E3}) is added by hand, which lacks a clear origin. Now we give a model considering the attractive Hubbard interaction, and the ground state calculated by the mean field approach can be BCS pairing state under appropriate parameters. 
\subsection{The Model and the Mean Field Approximation}
The model considering the attractive Hubbard interaction reads 
\begin{equation}
	H_{\rm{Hubbard}}=\sum_{\bm{k}} \psi^{\dagger}(\bm{k}) (H_{\rm{WSM}}(\bm{k})-\mu) \psi (\bm{k})-U\sum_{i}n_{i\uparrow}n_{i\downarrow},
\end{equation}
where $\psi(\bm{k})=(c_{\bm{k}, \uparrow}, c_{\bm{k}, \downarrow})^{T}$. We may perform the Fourier transform on the attractive Hubbard interaction term and obtain 
\begin{equation}
	\begin{aligned}
	H_{\rm{Hubbard}}&=\sum_{\bm{k}} \psi^{\dagger}(\bm{k}) (H_{\rm{WSM}}(\bm{k})-\mu) \psi (\bm{k})\\
	&-\frac{U}{V}\sum_{\bm{q}}\sum_{\bm{k}}\sum_{\bm{k'}} c^{\dagger}_{\bm{k'}+\bm{q},\uparrow}c^{\dagger}_{-\bm{k'},\downarrow}c_{-\bm{k},\downarrow}c_{\bm{k}+\bm{q},\uparrow}.
	\label{EA2}
	\end{aligned}
\end{equation}
As depicted in Fig. \ref{F4}(f), the Fermi surface for $\mu\neq0$ consists of two closed circles around the two Weyl points.  Thus, it is reasonable to consider the pairing with momentum $\bm{q}=\bm{0}$ and $\bm{q}=2\bm{Q}$ ($\bm{Q}=(0,0,\pi/2)$). Then we keep only two terms with  $\bm{q}=\bm{0}$ and $\bm{q}=2\bm{Q}$ in the summation of $\bm{q}$ in Eq. (\ref{EA2}). 
\begin{equation}
	\begin{aligned}
		H_{\rm{Hubbard}}&=\sum_{\bm{k}} \psi^{\dagger}(\bm{k}) (H_{\rm{WSM}}(\bm{k})-\mu) \psi (\bm{k})\\
		&-\frac{U}{V}\sum_{\bm{k}}\sum_{\bm{k'}} c^{\dagger}_{\bm{k'},\uparrow}c^{\dagger}_{-\bm{k'},\downarrow}c_{-\bm{k},\downarrow}c_{\bm{k},\uparrow}\\
		&-\frac{U}{V}\sum_{\bm{k}}\sum_{\bm{k'}} c^{\dagger}_{\bm{k'}+2\bm{Q},\uparrow}c^{\dagger}_{-\bm{k'},\downarrow}c_{-\bm{k},\downarrow}c_{\bm{k}+2\bm{Q},\uparrow}.
		\label{EA3}
	\end{aligned}
\end{equation}
Next we apply the mean field approximation and define the order parameter $\Delta_{\bm{q}}=-\frac{U}{V}\sum_{\bm{k}}\langle c_{-\bm{k},\downarrow}c_{\bm{k}+\bm{q},\uparrow}\rangle$. Correspondingly, the interaction term can be approximated as 
\begin{equation}
	\begin{aligned}
	&-\frac{U}{V}\sum_{\bm{k}}\sum_{\bm{k'}} c^{\dagger}_{\bm{k'}+\bm{q},\uparrow}c^{\dagger}_{-\bm{k'},\downarrow}c_{-\bm{k},\downarrow}c_{\bm{k}+\bm{q},\uparrow}\\
	&\approx \sum_{\bm{k}}\Delta^{*}_{\bm{q}}c_{-\bm{k}\downarrow}c_{\bm{k}+\bm{q},\uparrow}+\sum_{\bm{k}}\Delta_{\bm{q}}c^{\dagger}_{\bm{k}+\bm{q},\uparrow}c^{\dagger}_{-\bm{k},\downarrow}+\frac{V}{U}\vert\Delta_{\bm{q}}\vert^{2}. 
	\label{EA4}
	\end{aligned}
\end{equation}
Substituting the mean field approximation Eq. (\ref{EA4}) into Eq. (\ref{EA3}), we obtain the following result.
\begin{equation}
	\begin{aligned}
		&H_{\rm{Hubbard}}=\sum_{\bm{k}} \psi^{\dagger}(\bm{k}) (H_{\rm{WSM}}(\bm{k})-\mu) \psi (\bm{k})\\
		&+\sum_{\bm{k}}\Delta^{*}_{\bm{0}}c_{-\bm{k}\downarrow}c_{\bm{k},\uparrow}+\sum_{\bm{k}}\Delta_{\bm{0}} c^{\dagger}_{\bm{k},\uparrow}c^{\dagger}_{-\bm{k},\downarrow}+\frac{V}{U}\vert\Delta_{\bm{0}}\vert^{2}\\
		&+\sum_{\bm{k}}\Delta^{*}_{2\bm{Q}}c_{-\bm{k}\downarrow}c_{\bm{k}+2\bm{Q},\uparrow}+\sum_{\bm{k}}\Delta_{2\bm{Q}}c^{\dagger}_{\bm{k}+2\bm{Q},\uparrow}c^{\dagger}_{-\bm{k},\downarrow}+\frac{V}{U}\vert\Delta_{2\bm{Q}}\vert^{2}.
		\label{EA5}
	\end{aligned}
\end{equation}
To express the above model in the form of a BdG Hamiltonian, we rewrite each term in it.  First we deal with the normal state Hamiltonian of the Weyl semimetal. 
\begin{equation}
	\begin{aligned}
		&\sum_{\bm{k}} \psi^{\dagger}(\bm{k}) (H_{\rm{WSM}}(\bm{k})-\mu) \psi (\bm{k})\\
		&=\frac{1}{2}\{\sum_{\bm{k}} \psi^{\dagger}(\bm{k}) H_{N}(\bm{k}) \psi (\bm{k})+\sum_{\bm{k}} \psi^{\dagger}(\bm{k}) H_{N}(\bm{k}) \psi (\bm{k})\}\\
		&=\frac{1}{2}\sum_{\bm{k}} \psi^{\dagger}(\bm{k}) H_{N}(\bm{k}) \psi (\bm{k})+\frac{1}{2}\sum_{\bm{k}} \text{Tr} H_{N}(\bm{k})\\
		&+\frac{1}{2} \sum_{\bm{k}} -\psi^{T}(-\bm{k}) H^{*}_{N}(-\bm{k}) (\psi^{\dagger} (-\bm{k}))^{T}.
	\end{aligned}
\end{equation}
Here we have used the notation $H_{N}(\bm{k})=H_{\rm{WSM}}(\bm{k})-\mu$. Then the pairing terms are rewritten in the following form. 
\begin{equation}
	\begin{aligned}
    &\sum_{\bm{k}}\Delta^{*}_{\bm{q}}c_{-\bm{k}\downarrow}c_{\bm{k}+\bm{q},\uparrow}+\sum_{\bm{k}}\Delta_{\bm{q}}c^{\dagger}_{\bm{k}+\bm{q},\uparrow}c^{\dagger}_{-\bm{k},\downarrow}\\
    &=\frac{1}{2}\sum_{\bm{k}}\Delta^{*}_{\bm{q}}c_{-\bm{k}\downarrow}c_{\bm{k}+\bm{q},\uparrow}+\frac{1}{2}\sum_{\bm{k}}\Delta_{\bm{q}}c^{\dagger}_{\bm{k}+\bm{q},\uparrow}c^{\dagger}_{-\bm{k},\downarrow}\\
    &-\frac{1}{2}\sum_{\bm{k}}\Delta^{*}_{\bm{q}}c_{-\bm{k},\uparrow}c_{\bm{k}+\bm{q}\downarrow}-\frac{1}{2}\sum_{\bm{k}}\Delta_{\bm{q}}c^{\dagger}_{\bm{k}+\bm{q},\downarrow}c^{\dagger}_{-\bm{k},\uparrow}.
    \end{aligned}	
\end{equation}
Obviously, as mentioned above, we only need consider $\bm{q}=\bm{0}$ and $\bm{q}=2\bm{Q}$ cases.

\subsection{The Brillouin Zone Folding and the Self-consistent Equations}
Because the two pairing terms with  momenta $\bm{q}=\bm{0}$ and $\bm{q}=2\bm{Q}$ coexist in the Hamiltonian Eq. (\ref{EA5}), we cannot write it in the form of a BdG Hamiltonian under the basis $\Psi^{\dagger}(\bm{k})=(c^{\dagger}_{\bm{k}, \uparrow}, c^{\dagger}_{\bm{k}, \downarrow},c_{-\bm{k}, \uparrow}, c_{-\bm{k}, \downarrow})$ with $k_{x}$, $k_{y}$, and $k_{z}$ in the range $(-\pi,\pi]$. Instead, we have to consider the folded Brillouin zone (FBZ) with $k_{x}, k_{y}\in (-\pi,\pi]$ and $k_{z}\in (0,\pi]$, corresponding to the pairing momentum $2\bm{Q}=(0,0,\pi)$. Now the normal state Hamiltonian can be written as
\begin{equation}
	\begin{aligned}
	&\sum_{\bm{k}} \psi^{\dagger}(\bm{k}) H_{N}(\bm{k}) \psi (\bm{k})\\
	&=\sum_{\bm{k}\in \text{FBZ}} \psi^{\dagger}(\bm{k}) H_{N}(\bm{k}) \psi (\bm{k})\\
	&+\sum_{\bm{k}\in \text{FBZ}} \psi^{\dagger}(\bm{k}+2\bm{Q}) H_{N}(\bm{k}+2\bm{Q}) \psi (\bm{k}+2\bm{Q}).
	\end{aligned}
\end{equation}
Similarly, the pairing terms can also be written in the FBZ. 
Corresponding to the FBZ, the new basis becomes $\Psi^{\dagger}(\bm{k})=(\tilde{c}^{\dagger}_{\bm{k}},\tilde{c}^{\dagger}_{\bm{k}+2\bm{Q}},\tilde{c}_{-\bm{k}},\tilde{c}_{-\bm{k}-2\bm{Q}})$, where we have defined $\tilde{c}^{\dagger}_{\bm{k}}=(c^{\dagger}_{\bm{k},\uparrow}, c^{\dagger}_{\bm{k},\downarrow})$. Under the new basis considering the FBZ, the Hamiltonian now takes the following form
\begin{equation}
	\begin{aligned}
	H&=\frac{V}{U}\vert{\Delta_{\bm{0}}}\vert^{2}+\frac{V}{U}\vert{\Delta_{2\bm{Q}}}\vert^{2}+\frac{1}{2}\sum_{\bm{k}} \text{Tr} H_{N}(\bm{k})\\
	&+\frac{1}{2}\sum_{\bm{k}\in\text{FBZ}}\Psi^{\dagger}(\bm{k}) H_{\text{BdG}}(\bm{k})\Psi(\bm{k}),
	\end{aligned}
\end{equation}
where the explicit form of $H_{\text{BdG}}(\bm{k})$ is
\begin{equation}
\begin{pmatrix} 
	 H_{N}(\bm{k})& \bm{0}_{2\times2} & \tilde{\Delta}_{\bm{0}} &\tilde{\Delta}_{2\bm{Q}} \\ 
	 \bm{0}_{2\times2}& H_{N}(\bm{k}+2\bm{Q}) & \tilde{\Delta}_{2\bm{Q}}& \tilde{\Delta}_{\bm{0}}\\ 
	\tilde{\Delta}^{\dagger}_{\bm{0}} & \tilde{\Delta}^{\dagger}_{2\bm{Q}}& -H^{*}_{N}(-\bm{k})& \bm{0}_{2\times2}\\
	\tilde{\Delta}^{\dagger}_{2\bm{Q}}&\tilde{\Delta}^{\dagger}_{\bm{0}}& \bm{0}_{2\times2}& -H^{*}_{N}(-\bm{k}-2\bm{Q})
	\label{EA10}
\end{pmatrix}.
\end{equation}
Here we have used the notation
\begin{equation}
	\tilde{\Delta}_{\bm{q}}=\begin{pmatrix}
		0&\Delta_{\bm{q}}\\
		-\Delta_{\bm{q}}&0
	\end{pmatrix}.
\end{equation}

We suppose that we can diagonalize the BdG Hamiltonian in the following way
\begin{equation}
	\begin{aligned}
		&\frac{1}{2}\sum_{\bm{k}\in\text{FBZ}}\Psi^{\dagger}(\bm{k}) H_{\text{BdG}}(\bm{k})\Psi(\bm{k})\\
		=&\frac{1}{2}\sum_{\bm{k}\in\text{FBZ}}\Psi^{\dagger}(\bm{k})U(\bm{k})U^{-1}(\bm{k}) H_{\text{BdG}}(\bm{k})U(\bm{k})U^{-1}(\bm{k})\Psi(\bm{k})\\
		=&\frac{1}{2}\sum_{\bm{k}\in\text{FBZ}}\sum^{8}_{i=1} E_{i}(\bm{k})\Gamma^{\dagger}_{i}(\bm{k})\Gamma_{i}(\bm{k}), 
	\end{aligned}
\end{equation}
\begin{figure}[tbph]
	\centering
	\includegraphics[scale=0.344]{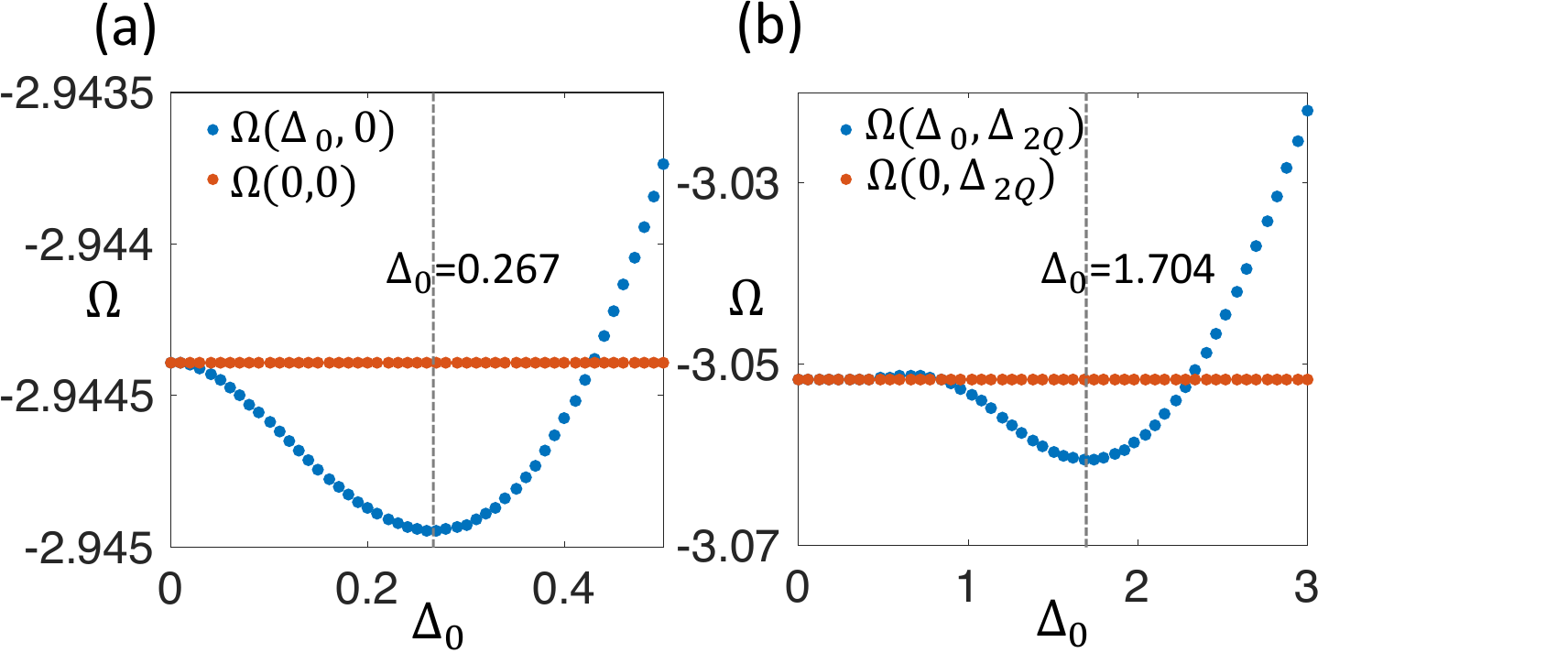}
	\caption{(a) The $\Omega-\Delta_{\bm{0}}$ relation with $\Delta_{2Q}=0$. The parameters are set as $\mu=0.6$ and $U=7$. (b) The  $\Omega-\Delta_{\bm{0}}$ relation with $\Delta_{\bm{2Q}}=0.741$. The parameters take the values $\mu=0.7$ and $U=8$.}
	\label{AF1}
\end{figure} 
in which the unitary matrix $U(\bm{k})$ diagonalizes $H_{\text{BdG}}(\bm{k})$, and we define $\Gamma(\bm{k})\equiv U^{-1}(\bm{k})\Psi(\bm{k})$. According to the definition of the order parameter $\Delta_{2\bm{Q}}=-\frac{U}{V}\sum_{\bm{k}}\langle c_{-\bm{k},\downarrow}c_{\bm{k}+2\bm{Q},\uparrow}\rangle$, we obtain the self-consistend equation
\begin{equation}
\begin{aligned}
	&\Delta_{2\bm{Q}}=-\frac{U}{V}\sum_{\bm{k}}\langle c_{-\bm{k},\downarrow}c_{\bm{k}+2\bm{Q},\uparrow}\rangle\\
	&=-\frac{U}{V}\sum_{\bm{k}\in \text{FBZ}}\langle c_{-\bm{k},\downarrow}c_{\bm{k}+2\bm{Q},\uparrow}\rangle-\frac{U}{V}\sum_{\bm{k}\in\text{FBZ}}\langle c_{-\bm{k}-2\bm{Q},\downarrow}c_{\bm{k},\uparrow}\rangle\\
	&=-\frac{U}{V}\sum_{\bm{k}\in \text{FBZ}}\langle \Gamma^{\dagger}_{n}U^{-1}_{n6} U_{3m}\Gamma_{m}\rangle-\frac{U}{V}\sum_{\bm{k}\in \text{FBZ}}\langle \Gamma^{\dagger}_{n}U^{-1}_{n8} U_{1m}\Gamma_{m}\rangle\\
	&=-\frac{U}{V}\sum_{\bm{k}\in \text{FBZ}}\delta_{mn}f(E^{n}_{\bm{k}})U^{-1}_{n6}U_{3m}+\delta_{mn}f(E^{n}_{\bm{k}})U^{-1}_{n8}U_{1m}.
	\label{EA13}
\end{aligned}
\end{equation}
Similarly, we can obtain the self-consistent equation of the order parameter $\Delta_{\bm{0}}$
\begin{equation}
	\begin{aligned}
		&\Delta_{\bm{0}}=-\frac{U}{V}\sum_{\bm{k}\in \text{FBZ}}\delta_{mn}f(E^{n}_{\bm{k}})U^{-1}_{n6}U_{1m}
		+\delta_{mn}f(E^{n}_{\bm{k}})U^{-1}_{n8}U_{3m}.
		\label{EA14}
	\end{aligned}
\end{equation}
In principle, we can obtain the order parameters $\Delta_{\bm{0}}$ and $\Delta_{2\bm{Q}}$ by solving the two self-consistent equations (\ref{EA13}) and (\ref{EA14}).

\subsection{The Free Energy}
The definition of free energy is $\Omega=-\frac{1}{\beta}\log Z$, where $Z$ is the partition function of the system we study. According to the knowledge in quantum statistics, we can calculate the partition function through the following formula
\begin{equation}
	Z=\text{Tr}e^{-\beta H}=e^{-\beta \text{Const.}}\text{Tr}e^{-\beta\frac{1}{2}\sum_{\bm{k}\in\text{FBZ}}\sum^{8}_{i=1} E_{i}(\bm{k})\Gamma^{\dagger}_{i}(\bm{k})\Gamma_{i}(\bm{k})}.
\end{equation}
 Here $\text{Const.}=\frac{V}{U}\vert{\Delta_{\bm{0}}}\vert^{2}+\frac{V}{U}\vert{\Delta_{2\bm{Q}}}\vert^{2}+\frac{1}{2}\sum_{\bm{k}} \text{Tr} H_{N}(\bm{k})$. We can readily evaluate the partition function \cite{PhysRevLett.128.037001}, because after 
 \begin{figure}[tbph]
	\centering
	\includegraphics[scale=0.375]{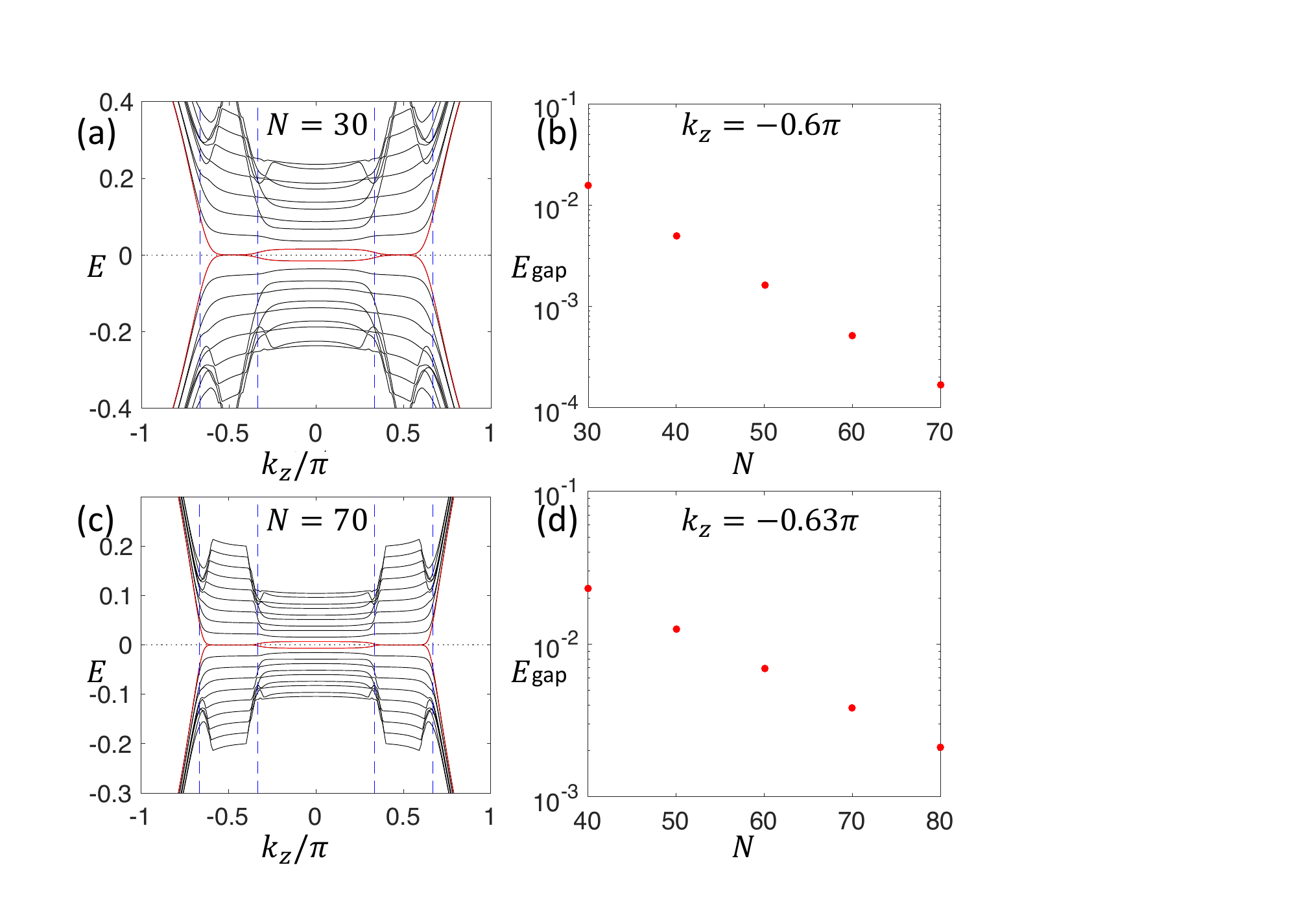}
	\caption{(a) and (c) show the VBSs of SC Weyl semimetals with $\mu=0$ and $\Delta=0.5$. The lattice sizes in (a) and (c) are $N=30$ and $N=70$, respectively. The exact phase boundaries are located at $k_{z}=\pm \pi/3$ and $\pm 2\pi/3$. (b) and (d) show how the gap of the VBSs (splitting of the two modes around $E=0$) changes as we increase the lattice size. (b) is calculated  at $k_{z}=-0.6\pi$, and (d) is calculated at $k_{z}=-0.63\pi$. Both (b) and (d) are calculated with the parameters $\mu=0$ and $\Delta=0.5$.}
	\label{AF2}
\end{figure} 
diagonalizing the BdG Hamiltonian, we obtain a system that consists of free fermions. Then we obtain the free energy
 \begin{equation}
 	\begin{aligned}
 	\Omega&=-\frac{1}{\beta}\log e^{-\beta \text{Const.}}\\
 	&-\frac{1}{2\beta}\log \prod_{\bm{k}\in \text{FBZ}}(1+e^{-\beta E^{1}_{\bm{k}}})\cdots(1+e^{-\beta E^{8}_{\bm{k}}})]\\
 	&=\frac{V}{U}\vert{\Delta_{\bm{0}}}\vert^{2}+\frac{V}{U}\vert{\Delta_{2\bm{Q}}}\vert^{2}+\frac{1}{2}\sum_{\bm{k}} \text{Tr} H_{N}(\bm{k})\\
 	&-\frac{1}{2\beta}\sum_{\bm{k}\in \text{FBZ}}\log (1+e^{-\beta E^{1}_{\bm{k}}})+\cdots +\log (1+e^{-\beta E^{8}_{\bm{k}}}).
 	\end{aligned}
 \end{equation} 

Since we may obtain the self-consistent solutions around the local minima of the free energy, we can compare the free energy of solutions around different local minima to obtain the the self-consistent solutions at the global minimum of the free energy. On the other hand, it is well known that the solutions of the self-consistent equations satisfying 
$\frac{\partial \Omega}{\partial \Delta}=0$. Thus, we may plot the $\Omega-\Delta$ relation to check whether we have solve the self-consistent equations correctly. Here is an example.  As shown in Fig. \ref{AF1}(a), when we set the parameters as 
$\mu=0.6$ and $U=7$, the self-consistent equations give the solutions $\Delta_{\bm{0}}=0.267$ and $\Delta_{\bm{2Q}}=0$, which is the BCS phase. On the other hand, since we have obtained the explicit expression for the free energy, we can plot the $\Omega-\Delta_{\bm{0}}$ relation with $\Delta_{\bm{2Q}}=0$. It is easy to find that the solution of the self-consistent equations $\Delta_{\bm{0}}=0.267$ minimizes the free energy. When the parameters take the values $\mu=0.7$ and $U=8$, the solutions of the self-consistent equations are $\Delta_{\bm{0}}=1.704$ and
$\Delta_{\bm{2Q}}=0.741$, corresponding to the coexistence of BCS phase and FF phase. Fig. {\ref{AF1}}(b) shows the $\Omega-\Delta_{\bm{0}}$ relation with $\Delta_{\bm{2Q}}=0.741$. Obviously, the solution $\Delta_{\bm{0}}=1.704$ minimizes the free energy, verifying our self-consistent calculations.  

\section{The Finite Size Effect around the Boundaries of the MFBs}
In the main text, we notice that although the real boundaries of the MFBs
are close to the exact boundaries given by the topological phase diagram, there is a notable difference between them. We have shown that this difference is due to the hybridization between the vortex MZM and the edge MZM.  In this section, we will illustrate that  increasing the lattice size can reduce this hybridization and extend the region of MFBs.

To be specific, we focus the $\mu=0$ and $\Delta=0.5$ case. Fig. \ref{AF2}(a) and Fig. \ref{AF2}(c)  show the VBSs of SC Weyl semimetals with lattice size $N=30$ and $N=70$, respectively. Obviously, compared to the regions of the MFBs in $N=30$ case, as we increase the lattice size to $N=70$, the regions of the MFBs are much more close to the exact phase boundaries (the blue dashed line). We notice that the MFBs occupy more space on the $k_{z}$ axis, but they are still enclosed by the exact boundaries (the blue dashed lines).

Then we observe the behavior of the gap between the two low-energy states in the VBSs at a fixed $k_{z}$ as we increase the lattice size. As shown in Fig. \ref{AF2}(b), we fix $k_{z}=-0.6\pi$ and increase the lattice size to calculate the VBSs. We find that the gap of the VBSs at $k_{z}=-0.6\pi$ decreases exponentially as the lattice size $N$ increases. A similar result appear in Fig. \ref{AF2}(d), in which the gap at $k_{z}=-0.63\pi$ also decreases exponentially as we increase $N$. Thus, we conclude that  the hybridization between the vortex core MZM and the edge MZM decreases as we increase the lattice size. Based on the results in Fig. \ref{AF2}, it is natural to expect that when the lattice size is large enough, the region of the MFBs will approach the exact phase boundaries. 

\bibliography{TRSB}
\bibliographystyle{apsrev4-2}
\end{document}